\documentclass[11pt]{article}
\usepackage[a4paper,margin=24mm]{geometry}
\usepackage[T1]{fontenc}
\usepackage{lmodern}
\usepackage{amsmath,amssymb,booktabs,tabularx,graphicx,placeins}
\usepackage{xcolor,listings,tikz}
\usetikzlibrary{arrows.meta,positioning}
\usepackage[hidelinks]{hyperref}
\definecolor{tigateal}{HTML}{087F76}
\definecolor{tigakeyword}{HTML}{174B8A}
\definecolor{tigastring}{HTML}{8C3B12}
\definecolor{tigacomment}{HTML}{52616B}
\definecolor{tigatype}{HTML}{087568}
\definecolor{tigacodebg}{HTML}{F7F9FC}
\lstdefinestyle{tigacode}{
  basicstyle=\small\ttfamily,
  keywordstyle=\color{tigakeyword}\bfseries,
  keywordstyle=[2]\color{tigatype},
  commentstyle=\color{tigacomment}\itshape,
  stringstyle=\color{tigastring},
  backgroundcolor=\color{tigacodebg},
  breaklines=true,columns=fullflexible,keepspaces=true,
  showstringspaces=false,upquote=true,tabsize=4,
  frame=single,rulecolor=\color{black!15},framesep=5pt,
  aboveskip=0.8em,belowskip=0.8em}
\lstdefinelanguage{TigaIR}{
  sensitive=true,
  alsoletter={._},
  morekeywords={module,func.func,return,arith.mulf,arith.addf,gf.apply,
    gf.yield,gf_iter.traverse,gf_kernel.launch,gf_storage.transfer,
    gf_storage.release,gf_storage.join},
  morekeywords=[2]{f16,f32,f64,i1,i32,i64,index,tensor,memref,
    block_rows,block_neighbors,num_warps,pipeline_stages},
  morecomment=[l]{//},
  morestring=[b]"}
\lstdefinelanguage{TigaShell}{
  sensitive=true,alsoletter={-},
  morekeywords={gf-opt},
  morekeywords=[2]{-gf-lower-domain-to-iter,-gf-lower-iter-to-kernel,
    -gf-select-kernel-schedule,-gf-plan-distributed-tasks},
  morecomment=[l]{\#},morestring=[b]",morestring=[b]'}
\newcommand{\code}[1]{\texttt{#1}}
\newcolumntype{Y}{>{\raggedright\arraybackslash}X}
\title{\includegraphics[width=38mm]{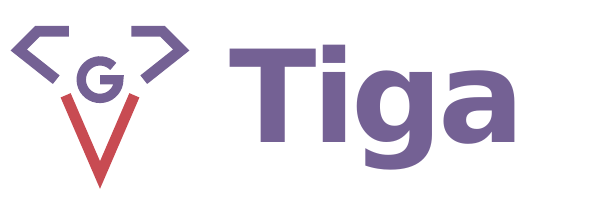}\\[6pt]
Tiga: Compiling Graph Message Passing at Scale}
\author{Mingyuan Chi\\Independent Researcher\\\href{mailto:walker.chi.000@gmail.com}{\texttt{walker.chi.000@gmail.com}}}
\date{Technical report --- September 21, 2026}
\hypersetup{pdfauthor={Mingyuan Chi},pdftitle={Tiga: Compiling Graph Message Passing at Scale}}
\begin{document}
\maketitle
\begin{abstract}
Graph message passing offers a common way to express learning algorithms,
physical simulations, and numerical solvers. Efficient execution depends on
interaction structure and data movement, which can be obscured when a program
is expressed as a sequence of tensor operations. On memory-constrained systems
such as laptops, materializing connectivity and intermediate messages can also
exhaust device memory. We present Tiga, a just-in-time compiler that separates
the definition of a message-passing program from how its interactions are
traversed, computed, and stored. Tiga preserves graph relations and reducer
algebra in a multi-level intermediate representation, enabling traversal
specialization, fusion of relation generation with aggregation, and reverse-mode
automatic differentiation. Backend-specific lowering targets CPUs and GPUs.
Its runtime extends execution beyond device-memory
capacity by streaming graph partitions from disk through host memory with
page-sized device staging buffers; partition ownership and halo exchange extend the
same programming model to distributed execution. A Python interface
interoperates with ordinary PyTorch tensors and autograd for forward and backward
computation. Numerical checks validate outputs and gradients for differentiable
workloads. A differentiated geometric workload characterizes the memory--time
tradeoff of fused forward and backward execution. Evaluation against matched Torch
and PyTorch Geometric baselines demonstrates reduced runtime and device-memory
use for generated-relation workloads, while offloaded forward execution
processes billion-edge graphs on a single memory-limited GPU. Measurements on
heterogeneous devices further characterize the communication and load-balance
costs of distributed execution.
\end{abstract}
\section{Introduction}
Message passing can be a general programming model, not merely an interface for
graph neural networks. Computation is organized as entities with state, relations
that determine which entities interact, local message functions, and successive
state updates. The graph need not be the input problem: it can represent the
structure of the computation itself. Tensor contractions, convolutions, particle
interactions, iterative solvers, attention and graph algorithms can all be viewed
through this lens. General message-passing models have a long history, including
the actor formalism~\cite{actors}.

We present Tiga (\emph{Target-Independent Graph Acceleration}), a just-in-time
(JIT) compiler for graph message-passing programs. Target independence refers to
the separation of program semantics from hardware-specific lowering: a message
function describes an interaction, while traversal, kernel mapping and data
placement determine its execution. The implementation targets CPU and NVIDIA
GPU backends, with an extensible backend integration interface for additional targets.

The long-term goal is a common programming substrate for dense, sparse and
dynamically generated computation, rather than a collection of unrelated
operator-specific interfaces. Relations describe interaction, messages describe
local computation, reducers describe combination, and composition with control
flow describes how state evolves. A compiler can then choose how these programs
execute across kernels, memory tiers and devices while preserving their meaning.
Generality concerns what can be expressed; it does not imply that every problem
admits an efficient parallel encoding or that every encoding is supported by the
current implementation. Section~\ref{sec:general-message-passing} makes this
distinction precise.

This common form does not imply a common optimal implementation. A relation in
compressed sparse row (CSR) storage needs an indexed traversal; a dense relation may need no edge array;
a generated radius relation may profit from a spatial directory. Feature width,
degree skew, gradient requirements and data residency change the suitable plan.

Three practical costs motivate this design. First, explicitly expanding a
radius or nearest-neighbor relation can create indices, geometry and messages
that are consumed only once. Second, a graph larger than GPU memory requires
bounded storage movement even when its arithmetic is simple. Third, dividing
a graph across devices introduces halo traffic and rank imbalance; doubling
the GPU count need not reduce completion time.

The central idea in Tiga is to separate logical interactions from their physical
execution. Keeping relations visible to the compiler allows generated neighbors
to be consumed directly by aggregation instead of first being stored as a full
edge list. Separating logical values from their residence allows a large graph
to be processed in pages, while ownership and halo exchange define its
distributed execution. Backend-specific lowering then maps the computation to
CPU or GPU kernels. These are complementary ways to control the cost of the
same program, rather than separate application models. For memory-constrained
systems such as laptops, paging is particularly important: graph capacity need
not be limited to accelerator memory. Current executable backends and extension
boundaries are distinguished in Table~\ref{tab:backends}.

Tiga retains relations and reducer algebra as compiler-visible information.
It is not a model library, optimizer package or replacement for PyTorch.
Ordinary applications retain \code{torch.Tensor} and PyTorch
autograd~\cite{pytorch}. Reverse-mode automatic differentiation computes
vector--Jacobian products (VJPs), propagating an output loss gradient back to
input fields and model parameters~\cite{autodiff}. Torch is
not a base installation dependency: a standalone native path supplies CPU
computation, message passing and gradients without it. Native tensors also
expose compiler/storage functionality not uniformly available through Torch.
This split is an explicit interoperability boundary, not a claim that both
interfaces support all operations identically.

The report makes the following contributions:
\begin{itemize}
\item A relation-aware compiler design that carries field roles, reducer algebra
and derivative requirements from the programming interface to traversal and
kernel selection.
\item A runtime model that separates logical values from storage residence and
rank ownership, enabling paged and distributed execution with explicit data
movement contracts.
\item An evaluation of runtime and memory efficiency, billion-edge capacity,
and heterogeneous distributed execution, accompanied by numerical checks,
reproducible measurements, and a differentiable EdgeNN example.
\end{itemize}
The contribution is the integration of these mechanisms around a shared
message-passing representation; Section~\ref{sec:background} relates the
individual ideas to prior work.
The experiments compare time and allocation with matched Torch/PyG paths,
execute one billion explicit edges on a 16 GiB GPU, and measure the cost of
two-host NCCL graph partitioning. The results show that
generated radius traversal and exact kNN improve latency and allocation on the
measured workloads; paging extends executable capacity; and the current
equal-row two-GPU schedule is slower than the faster GPU alone. This connects
each mechanism to a measured benefit or cost rather than assuming that one
execution strategy is efficient for every relation and interconnect.

\paragraph{Reading and using the system.}
Application development starts with the programming model in
Section~\ref{sec:programming-model} and the complete forward/backward example in
Appendix~\ref{app:edge-nn}; Appendix~\ref{app:attention} expresses causal attention
with the same graph, edge and reducer interface. Compiler and runtime development is covered in
Sections~\ref{sec:compiler}--\ref{sec:runtime}; reproduction details appear in
Section~\ref{sec:evaluation}. The
\href{https://github.com/walkerchi/TIGA-lang}{implementation},
\href{https://walkerchi.github.io/TIGA-lang/}{API documentation}, and
\href{https://github.com/walkerchi/tiga-lang-paper}{report artifacts}
are available separately so that implementation details and experiments can
be inspected alongside the design.

\section{Background: why graph-aware compilation?}
\label{sec:background}
\subsection{Message passing as a general computational model}
\label{sec:general-message-passing}
Message passing is broader than neighborhood aggregation in a fixed-depth
graph neural network (GNN).
In the general model, stateful entities perform local computation, communicate
values or control information, and continue through repeated transitions.
The actor formalism treats communication as an organizing principle for data,
procedures and control~\cite{actors}. Pregel gives a graph-oriented, iterative
realization through vertex programs and messages~\cite{pregel}. Tiga uses
structured relations and reducers rather than implementing arbitrary actor
mailboxes or adopting Pregel's execution protocol.
In graph learning, Message Passing Neural Networks (MPNNs) formalize learned
messages and aggregation across repeated node updates~\cite{mpnn}. Tiga uses
this local-computation interface as a compiler input, while also supporting
non-neural message functions such as stencil updates and weighted sums.

\paragraph{Universality and its conditions.}
With sufficiently expressive local transitions, repeatable control and
arbitrarily extensible state, a message-passing model can in principle express
any computable algorithm. A constructive way to see this is to encode a Turing
machine's tape as a chain of entities. Each entity stores a tape symbol; a
message carries the head's finite control state. Receiving the head message
updates the symbol and sends the next head state to the left or right neighbor.
Allowing the chain and the number of transitions to grow reproduces the machine's
computation. This is an expressiveness argument for the general model, not a
universality proof for Tiga's present bounded loops, fixed-shape tensors or
finite-memory backends. It refers to computable problems, not undecidable ones,
and says nothing by itself about time, memory or useful parallelism.

\paragraph{A common language for numerical programs.}
The practical opportunity is to retain a useful computational structure without
requiring every application to start from an explicit graph dataset. Elementwise
maps use independent entities or self-relations. Matrix multiplication connects
each output row to contributing source rows, with weighted vector messages and
a sum reducer. Convolution and finite-difference stencils use regular local
relations; particles use coordinate-dependent radius or kNN relations; attention
uses dense or masked relations and a normalization-aware reducer. Repeated
message passing with state updates expresses time stepping, iterative solvers
and graph traversals. These are representations within a shared conceptual model,
not a claim that every specialized operation has an equally optimized Tiga path.
The language ambition is to make such programs composable and their interactions
compiler-visible; the implementation and evaluation below establish concrete
slices of that ambition.

\paragraph{Why a compiler is needed.}
An edge function can be only a multiply, yet its execution may require reading
an index, gathering a feature vector, creating an edge-sized temporary and
scattering into a destination. For $E$ edges and feature width $F$, explicitly
materializing messages requires storage proportional to $EF$, even when the
final output has only $NF$ values. A fused traversal can instead accumulate
messages directly. Fusion is not universally optimal: a library sparse matrix
operation, feature tiling, recomputation or a different graph layout may win
depending on shape, degree distribution and reuse. Consequently the relevant
compiler question is which semantic information remains available when each
decision is made, not whether a system uses the phrase ``message passing.''

PyTorch Geometric supplies a graph-learning programming model with gather--scatter
and optimized sparse execution paths~\cite{pyg,pyg2}. Graphiler makes the compiler connection explicit:
its message-passing data-flow graph annotates graph-related tensor residency and
movement, enabling reductions in redundant computation and intermediate
storage~\cite{graphiler}. Tiga likewise retains source-, edge- and
destination-associated semantics, and carries these roles into relation traversal
and differentiation.

\subsection{Graph and feature schedules, fusion, and training}
FeatGraph composes sparse traversal templates with feature-dimension UDFs and
schedules~\cite{featgraph}. It shows why graph traversal and within-feature
computation should be optimized together. Seastar offers vertex-centric Python
programming and generates fused forward and backward GPU kernels~\cite{seastar}.
Tiga shares the goal of fusing local computations, and combines it with
explicit relation constructors and storage/ownership representations.

GALA is especially relevant to a multi-level design. It separates data and
compute representations, composes intra-operator and inter-operator
transformations, and reasons about training-specific optimization opportunities
across forward and backward computations~\cite{gala}. Tiga separates relation,
traversal and execution representations; its current evaluation focuses on
forward workloads, with path-specific gradient implementations described in
the differentiation section. Whole-training-program optimization remains beyond
that evaluated scope.

GraphIt separates the graph algorithm from scheduling choices~\cite{graphit}.
That distinction helps explain why a relation-level operation should not
prematurely encode threads, blocks or traversal order. Tiga additionally seeks
to preserve numerical field operations, reducer state and differentiation
requirements alongside traversal scheduling.

\subsection{Relations and sparse iteration beyond GNNs}
Ebb represents geometric domains through a relational data model and separates
simulation code, domain libraries and parallel runtime implementation~\cite{ebb}.
Simit connects hypergraph-local computation to global tensor algebra for physical
simulation~\cite{simit}. They are important precedents for a graph language that
is not limited to neural networks. Tiga's particle, solver and stencil examples
belong in this lineage of relational numerical programming.

TACO derives compound dense/sparse tensor kernels from high-level algebra using
iteration and merge structures~\cite{taco}. Its format-abstraction work further
separates logical computation from sparse storage levels~\cite{sparseformats}.
SparseTIR develops composable sparse formats and transformations for deep-learning
operators~\cite{sparsetir}. These works motivate keeping coordinate traversal
distinct from physical layout and execution mapping. Tiga's Iter IR preserves
relation-coordinate structure for its supported traversal families; the
implementation does not provide a general sparse-format algebra.

Triton supplies a lower-level tiled compilation interface~\cite{triton}, while
MLIR supplies infrastructure for composing dialects and lowering passes~\cite{mlir}.
Tiga uses both: graph-aware representations are upstream of a current NVIDIA
provider. The provider performs device-code compilation after Tiga's graph-aware
transformations.

\subsection{Capacity: running the graph at all}
Reducing execution latency and reducing the resident working set are related but
distinct goals. If a graph does not fit a resource budget, a faster in-memory
kernel is insufficient. GraphChi established disk-based large-graph computation
on a single machine through partitioned processing and sliding
windows~\cite{graphchi}. X-Stream uses edge-centric streaming partitions to favor
sequential access over random graph-data access~\cite{xstream}. These systems are
relevant even though they do not share Tiga's differentiable Tensor interface.

Legion separates logical regions and task privileges from placement and physical
instances, providing a model for reasoning about locality, independence and
data movement~\cite{legion}. Tiga's storage versions, task dependencies and halo
ownership apply this separation to the graph runtime described in Section~\ref{sec:runtime}.

These systems distinguish executing faster from making execution possible.
Tiga evaluates both, plus the cost of distributing the same graph. The capacity
experiment traverses explicit disk-backed CSR through host staging to CUDA;
the distributed experiment measures rank ownership and halo movement separately.
Neither a transfer-bandwidth microbenchmark nor a successful communicator setup
alone answers these application-level questions.

FlashAttention is a related example of IO-aware exact
computation~\cite{flashattention}: avoiding an intermediate can matter as much as
reducing arithmetic. Tiga's dense relations and streaming reducers expose
similar scheduling opportunities, while optional tile pruning changes the
mathematical contract and is distinct from exact attention. The experiments in
this report concern graph aggregation and do not claim attention-library parity.

\begin{table}[ht]
\centering\small
\begin{tabularx}{\linewidth}{l Y Y}
\toprule
Prior work & Relevant mechanism & Connection to Tiga \\
\midrule
Graphiler~\cite{graphiler} & Graph-aware dataflow/residency & Source, edge and destination field roles \\
FeatGraph~\cite{featgraph} & Sparse traversal + feature schedules & Joint graph/feature traversal selection \\
Seastar~\cite{seastar} & Vertex-centric fused forward/backward & Message fusion and derivative generation \\
GALA~\cite{gala} & Composed data/compute and training transforms & Layered representation and transformation scope \\
Ebb / Simit~\cite{ebb,simit} & Relational simulation / hypergraph algebra & Numerical programs beyond GNNs \\
GraphIt~\cite{graphit} & Algorithm/schedule separation & Separate meaning from execution mapping \\
TACO / SparseTIR~\cite{taco,sparsetir} & Sparse iteration, formats, transformations & Relation traversal separate from storage \\
GraphChi / X-Stream~\cite{graphchi,xstream} & Disk-oriented partitioned execution & Bounded graph pages and storage traffic \\
Legion~\cite{legion} & Logical regions and physical placement & Storage instances, ownership and dependencies \\
\bottomrule
\end{tabularx}
\caption{Prior work and its relationship to Tiga's design.}
\end{table}
\paragraph{Scope of the distinction.}
Tiga does not introduce message passing, kernel fusion, sparse traversal or
offloading individually. Its design contribution is their integration around
explicit relation semantics: a generated relation can feed a reduction without
a stored edge list, while a stored relation can retain the same local program
under a bounded-residency runtime. Field roles and reducer contracts constrain
legal transformations, and differentiation is attached to the supported
execution route. The distinction from a library of operators is the retained
program representation; the distinction from existing compilers is the
particular combination of relation constructors and storage/ownership contracts,
not a claim of uniformly broader coverage or faster execution. Measurements
below compare implemented paths, not every cited system.
\FloatBarrier

\section{Programming model}
\label{sec:programming-model}
\subsection{Relations, fields and messages}
Let source entities and destination entities be distinct finite sets. A relation
is an ordered collection of edges with source and destination maps. Duplicate
edges remain separate messages. For a destination, the computation is
$$
 m_e = f(x_{s(e)}, z_{d(e)}, a_e;\theta),\qquad
 y_i = g\!\left(z_i,\operatorname{finalize}\left(
       \bigoplus_{e:d(e)=i}\operatorname{lift}(m_e)\right);\theta\right).
$$
Here source fields, destination fields, edge attributes and shared parameters
are denoted by $x$, $z$, $a$ and $\theta$, respectively. The edge function $f$
forms a message; $g$ optionally updates destination state from the aggregate.
The reducer supplies its identity, lift, combine and finalize behavior.
An empty row starts from the identity; finalization determines the resulting
empty-row convention. This convention is reducer-specific. Declaring a reducer
does not define which edge-local values enter it: the edge function does.
Simple sum messages return a tensor directly. A structured streaming reducer
such as online softmax currently packages a score and value using an explicit
reducer call inside the edge function. That call does not execute an aggregation
once per edge; it binds the message inputs for capture.

CSR is destination-oriented: row boundaries enumerate destinations and column
indices identify sources. The example below has two edges into destination 0,
one into destination 1, and none into destination 2. A self edge is ordinary;
there is no automatic reverse edge.
\noindent\begin{minipage}{\linewidth}
\begin{lstlisting}[language=Python]
import torch
import tiga as tg

class WeightedSum(tg.MessagePassing):
    reducer = tg.sum()
    def edge(self, src, dst, edge):
        return src.x * edge.w

graph = tg.Graph.from_csr(
    torch.tensor([0, 2, 3, 3]), torch.tensor([0, 1, 1]),
    num_src=2, validate="full")
x = torch.tensor([2., 3.], requires_grad=True)
w = torch.tensor([4., 5., 2.], requires_grad=True)
y = WeightedSum()(graph=graph, src={"x": x}, dst={}, edge={"w": w})
dx, dw = torch.autograd.grad(y.sum(), (x, w))
# y=[23,6,0], dx=[4,7], dw=[2,3,3]
\end{lstlisting}
\end{minipage}

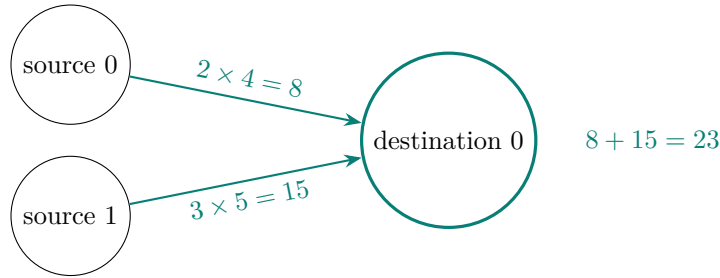
\begin{figure}[ht]
\centering
\begin{tikzpicture}[>=Stealth, every node/.style={font=\small}]
\node[draw,circle] (s0) at (0,1.0) {source 0};
\node[draw,circle] (s1) at (0,-1.0) {source 1};
\node[draw,circle,very thick,draw=tigateal] (d0) at (5,0) {destination 0};
\draw[->,thick,color=tigateal] (s0) -- node[above,sloped] {$2\times4=8$} (d0);
\draw[->,thick,color=tigateal] (s1) -- node[below,sloped] {$3\times5=15$} (d0);
\node[right=0.5cm of d0,text=tigateal] {$8+15=23$};
\end{tikzpicture}
\caption{One destination row: two edge messages become one result. Other
destinations are intentionally omitted.}
\end{figure}

\subsection{Composition and topology}
Dense and triangular constructors describe relation structure, while
\code{Graph.cat} forms disjoint blocks by offsetting source and destination
IDs independently. For example, concatenating triangular graphs of sizes two
and three represents independent causal sequences. It introduces no cross-block
edges and currently materializes CSR; generic composition is not yet an
implicit-storage optimization. Cumulative sequence boundaries remain a
compatibility constructor and must start at zero.

Radius and nearest-neighbor builders depend on coordinates. Their relation
realization may differ by execution path. Topology selection is discrete;
the selected graph is a fixed snapshot for derivative semantics. In-place
mutation of captured indices is not a supported topology-update mechanism.
Transpose and composition can reorder edges, so explicit edge fields require
matching reordering. A constructor accepting an input does not certify every
consumer, provider or gradient combination.

\section{Compiler organization}
\label{sec:compiler}
Tiga uses the extensible Multi-Level Intermediate Representation (MLIR)
infrastructure~\cite{mlir}. An intermediate representation (IR) records a program
in a form that compiler passes can analyze and transform. Tiga's representations separate
the mathematical computation from traversal and execution decisions.
Applications use the Python interface; the \code{gf.*} dialect operations below
are its compiler-facing representation. Domain, Iter, Kernel and Task/Storage
name semantic layers, each containing several operations.

\begin{table}[ht]
\centering\small
\begin{tabularx}{\linewidth}{l l Y}
\toprule
Layer & Example operation & Preserved decision \\
\midrule
Domain & \code{gf.apply} & Relation, field roles, message regions and reducer algebra \\
Iter & \code{gf\_iter.traverse} & Traversal organization and reduction structure \\
Kernel & \code{gf\_kernel.launch} & Work mapping and local resource organization \\
Task/Storage & \code{gf\_task.launch} & Dependency, transfer and readiness ordering \\
Tensor & \code{gf\_tensor.mul} & Tensor expressions, views and gradient computation \\
\bottomrule
\end{tabularx}
\caption{Compiler representations and the decisions they preserve. Execution
paths use the layers needed by the selected operation and backend.}
\end{table}

\subsection{Design contract: delay decisions, retain legality}
\label{sec:ir-design}
The layers answer different questions about the same program. Domain asks what
messages and reductions mean. Iter asks which relation coordinates must be
visited. Kernel asks how that work is mapped to execution resources. Task and
Storage ask when work may run and which physical copy supplies its operands.
Tensor IR is a numerical-expression layer used alongside these decisions, not
necessarily a final fifth stage. The split avoids encoding a GPU schedule in
the mathematical interface, but does not make scheduling automatic or optimal.

The key invariant is preservation of the observable outputs and supported VJPs
for the declared input contract. Shape, dtype, device, field role, reducer state,
topology snapshot and data-access effects are parts of that contract. A pass
that cannot establish a required property must reject the transformation or
select an explicitly diagnosed alternative. A topology version in the IR is a
validity annotation; it does not make arbitrary external in-place mutation safe.

\subsection{Domain: field roles and reducer algebra}
In the example, the source field is indexed by a source ID and the edge weight
by an edge position. Both are floating-point vectors, but exchanging their roles
changes the program. \code{gf.apply} therefore retains \code{input\_roles},
\code{input\_names}, reducer symbols, region kinds and input segments. Its region
contains the edge-local arithmetic, not an already materialized global message
array. A relation separately retains cardinalities, origin, lifecycle, version
and the available degree constraints.

A reducer describes identity, lift, combine and finalize regions, including the
types of its intermediate state. This is richer than attaching the string
``sum'' to an arbitrary scatter. A product or online-softmax reducer needs
different state and different differentiation rules. Empty rows start from the
identity and still obey finalize. Associativity admits regrouping; commutativity
admits reordering. These declarations are semantic obligations, not properties
the compiler proves for unrestricted user-written Python. Floating-point sums
require tolerance-aware validation when their reduction tree changes.

The full fixture and four unedited compiler outputs are included in
\code{generated/ir/}. This excerpt shows the fixture's edge region; its
enclosing Domain operation is omitted:
\begin{lstlisting}[language=TigaIR]
^bb0(%source: f32, %edge: f32):
  %message = arith.mulf %source, %edge : f32
  "gf.yield"(%message) : (f32) -> ()
\end{lstlisting}
The block arguments represent one logical edge invocation. The region itself
does not say which CPU core, CUDA thread or memory tier executes it.

This representation makes the materialization decision explicit. An eager
implementation gathers a source row for every edge, writes its weighted
message and then reduces those messages. A fused row traversal reads the same
source and weight, forms a tile-local message and immediately updates reducer
state. For feature width $F$, the former can allocate an edge-message array
with $EF$ entries; the latter needs only tile state in addition to graph,
inputs and the $NF$ output. The edge set and message arithmetic are unchanged.
The fixed-topology study in Section~\ref{sec:supplement} holds that edge set
constant to examine execution-path differences. Generated radius traversal
additionally avoids accepted-edge storage, but that distinct benefit is not
isolated by a stored-CSR comparison.

\subsection{Iter: coordinates before threads}
The Domain-to-Iter pass replaces the application with \code{gf\_iter.traverse}
and preserves the reducer and edge region. In the checked CSR fixture, the
attributes \code{coordinate\_hierarchy = "compressed-row"} and
\code{ordering = "destination-major"} describe a row-oriented coordinate walk.
These names do not mean that the operation is already an explicit sequential
loop, or that all threads use the same row length.

For the three-row example, \code{row\_ptr=[0,2,3,3]} determines edge ranges
$[0,2)$, $[2,3)$ and $[3,3)$. The edge position indexes the weight;
\code{col\_idx[e]} supplies the source ID. Thus the first row visits source IDs
0 and 1, forms messages 8 and 15 and produces 23. The second produces 6; the
empty third row produces the sum identity 0. Confusing edge positions with
source IDs would already invalidate lowering before any hardware schedule.

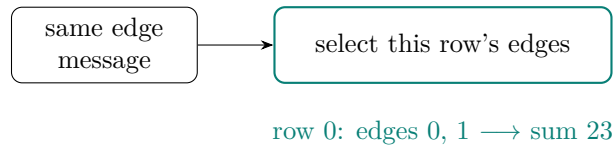
\begin{figure}[htbp]
\centering
\begin{tikzpicture}[>=Stealth,every node/.style={font=\small}]
\node[draw,rounded corners,text width=2.2cm,align=center,minimum height=1cm] (domain) {same edge message};
\node[draw,rounded corners,draw=tigateal,thick,text width=4.2cm,align=center,minimum height=1cm,right=1cm of domain] (iter) {select this row's edges};
\draw[->] (domain) -- (iter);
\node[below=0.35cm of iter,text=tigateal] {row 0: edges 0, 1 $\longrightarrow$ sum 23};
\end{tikzpicture}
\caption{Iter introduces coordinate traversal, not a thread assignment.
The highlighted decision is row membership.}
\label{fig:iter-design}
\end{figure}

Dense and generated relations need different coordinate producers. Materializing
all edges merely to resemble CSR can destroy the reason to retain an implicit
relation. Conversely, preserving a constructor in the frontend does not prove
that every backend consumes it implicitly: graph concatenation currently
materializes CSR. Format abstraction and schedule selection must be inspected
on the path actually chosen.

\subsection{Kernel: a concrete, guarded schedule}
The fixture's Kernel operation consumes CSR buffers and field tensors. Its
selected bounded-ragged schedule records these attributes:
\begin{lstlisting}[language=TigaIR]
block_rows = 16, block_neighbors = 16,
num_warps = 4, pipeline_stages = 1
\end{lstlisting}
These are values for this fixture, not global Tiga defaults.
Its execution roles distinguish workgroup destination rows, subgroup
neighbor reduction and scalar lanes. Resource and handoff attributes describe
global reads, private register state, subgroup reduction and global output writes.

The important transition is from enumerating the correct coordinates to deciding
how multiple coordinates share hardware. A maximum-degree guard can admit a
masked bounded-neighbor tile; without that bound, the same launch geometry may
be incorrect or wasteful. A high-degree split requires explicit partial states
and a legal merge. No universal schedule should be inferred from the simple
case's kernel. Provider-neutral scheduling still needs target capabilities and
resource constraints; it does not imply identical machine instructions on every
backend.

\subsection{Task and Storage: dependencies are not residency}
The checked distributed fixture introduces partition and halo planning, then
pack, exchange and unpack events. Interior work depends on partition readiness;
boundary work additionally depends on unpack completion. A join represents the
completion of both branches. This fixture demonstrates the dependency vocabulary;
the public executor uses the communication-then-compute order described in
Section~\ref{sec:runtime}.

Storage represents a logical region and version separately from physical
instances, which carry capacity, memory space, device and layout.
\code{gf\_storage.transfer} produces a readiness event;
\code{gf\_storage.release} depends on completion before storage is recycled;
\code{gf\_storage.join} joins events without specifying a host-wide synchronize.
These operations establish a vocabulary for movement and lifetime, not proof
that a cost model automatically supplies an out-of-core schedule for any kernel.

For a paged graph, correctness requires that the page's topology and field
version agree, transfer completes before consumption, and storage is not reused
until the last consumer finishes. The current public CPU/CUDA paged-CSR executor
uses host-side orchestration for these dependencies. Whole-Tensor
least-recently-used (LRU) eviction is a
separate policy from partitioning one oversized active working set.

\subsection{Transformations, verification and differentiation}
\begin{table}[ht]
\centering\small
\begin{tabularx}{\linewidth}{l Y Y}
\toprule
Transformation & Required fact & Failure if ignored \\
\midrule
Row regrouping & Reducer association/order contract & Different state or numerical behavior \\
Bounded neighbor tile & Degree/shape guards and masked accesses & Truncated rows or out-of-range loads \\
Producer--consumer fusion & Compatible effects and dependencies & Stale values or duplicated side effects \\
Topology reuse & Matching frozen snapshot and bindings & Wrong edges after mutation \\
Storage reuse & Last consumer completed & Use-after-recycle or stale version \\
VJP construction & Saved/recomputed primal state and fixed topology & Incorrect input/parameter gradients \\
\bottomrule
\end{tabularx}
\caption{Legality obligations, not a claim that every possible violation is
statically proven absent. Runtime guards and negative tests remain necessary.}
\end{table}

Differentiation can change the access pattern: source gradients accumulate over
reverse incidence, and edge gradients retain edge identity. A forward schedule
need not be a good backward schedule. The compiler therefore needs a distinction
between semantic VJP generation, schedule selection and runtime saved-state
management. For current Torch CSR replay, that semantic bridge is implemented
through Torch autograd rather than a generated TTIR backward kernel; dedicated
edge-NN and native VJP paths have different contracts.

From the report repository, reproduce the architectural example with an
installed \code{gf-opt} compiler executable:
\begin{lstlisting}[language=TigaShell]
gf-opt generated/ir/distributed-plan.mlir \
  -gf-lower-domain-to-iter \
  -gf-lower-iter-to-kernel -gf-select-kernel-schedule \
  -gf-plan-distributed-tasks
\end{lstlisting}
The \path{generated/ir/} directory contains the individual stages in
\code{domain.mlir}, \code{iter.mlir}, \code{kernel.mlir} and \code{task.mlir}.
The artifact exporter checks them against compiler output before copying them.

\subsection{Implemented routes through the layers}
The source example \code{generated/ir/distributed-plan.mlir} contains explicit
placement metadata and follows the Domain, Iter, Kernel and Task progression.
It is supplied as a compiler-level input so that each lowering decision can be
reproduced independently of Python capture.

Native Tensor programs, including native CSR message passing, can lower through
Tensor IR to Vector/SCF/MemRef and LLVM execution. Relation-oriented CUDA paths
retain graph structure through Iter and Kernel representations. Task and Storage
operations appear where needed for dependencies and movement; a simple call does
not necessarily materialize each architectural layer as an inspectable artifact.

\subsection{Providers and inspectability}
A backend is an execution target and its lowering route; a provider is the
integration component that invokes a target toolchain and exposes its executable
to the runtime. A transport moves data between ranks and does not generate
compute kernels.

The NVIDIA provider hands serialized Triton IR (TTIR) to Triton~\cite{triton}, which
continues to device code. Serialization avoids requiring the provider and Tiga
to share one in-process MLIR ABI. Target capability information still influences
earlier scheduling. CPU execution uses LLVM and an ExecutionEngine binding.
Table~\ref{tab:backends} distinguishes executable routes from extension interfaces.

\begin{table}[ht]
\centering\small
\begin{tabularx}{\linewidth}{l Y Y}
\toprule
Backend / target & Current executable route & Evidence and boundary \\
\midrule
CPU / LLVM & Tensor IR through Vector/SCF/MemRef and LLVM ExecutionEngine & Native CPU forward/VJP and paged CSR; supported reducer/layout subset \\
NVIDIA / CUDA & Relation/Kernel IR to serialized TTIR, vendor Triton, PTX/cubin; CUDA Driver runtime & Torch/PyG comparisons, billion-edge paged forward and two-host NCCL execution; backward remains path-dependent \\
Torch adapter & Ordinary Torch tensors and autograd; dispatch to compiled or library/reference routes & Interoperability layer, not an independent hardware backend; some CSR VJPs replay Torch \\
AMD / Hygon & Provider ABI and conformance scaffolding & Pending provider integration and real-hardware validation; no speedup reported \\
Apple Metal & Provider interface only & Pending legal lowering and device validation; no speedup reported \\
Other accelerators & Extension points & Device implementations require provider integration and hardware validation \\
\bottomrule
\end{tabularx}
\caption{Backend execution paths. CUDA graph paging uses host staging,
and MPI/TCP/NCCL are transports rather than kernel code generators.}
\label{tab:backends}
\end{table}

The application can inspect \code{program.explain()} and available stage/code
artifacts. Native tensors additionally expose execution metadata after
realization. Available artifacts depend on the selected provider.
Small native expressions under an automatic policy may use a
Python reference evaluator. Strict native validation explicitly requires the
native policy and checks the reported backend.

For a new target, the implementation work lies in capability checks, legal
lowering, provider compilation/loading, and runtime buffer and event handling.
The common IR preserves the computation, while those components account for
target-specific layouts and resource limits. The
\href{https://github.com/walkerchi/TIGA-lang/blob/main/python/tiga/codegen/registry.py}{provider interface}
and \href{https://walkerchi.github.io/TIGA-lang/development/}{development guide}
provide implementation entry points. Forward and VJP conformance must be checked
separately; adding a device name alone does not supply either executable path.

\subsection{Generated relations and executable reuse}
Generated relations allow a consumer to evaluate candidate neighbors and
accumulate messages without storing the accepted edges or messages globally.
For low-dimensional Euclidean radius relations, a cell directory restricts
candidate traversal to nearby spatial cells. An exact distance predicate then
selects the accepted neighbors. The benefit depends on cell occupancy and
geometry; a custom distance function requires a compatible spatial bound before
this traversal is legal. Section~\ref{sec:performance} evaluates a radius workload
that uses this build-and-consume strategy.

For generated kNN, one program scans candidate tiles for a destination query,
retains a power-of-two padded top-k state, and consumes selected source features
without materializing adjacency. A candidate's ordering key contains its squared
distance and source index, so distance ties have a stable order. Before sorting
a tile, the kernel compares its minimum key with the largest retained key.
If the tile cannot improve the retained state, sorting and merging are skipped
exactly; distance evaluation is not skipped. Otherwise, local sorting and the
retained-state merge execute normally. For non-power-of-two k, the padded state
makes the test conservative. The search still examines all query--candidate
pairs and is not a spatial-index algorithm.

Repeated public kNN calls can bind coordinates from a new Graph descriptor with
equivalent relation semantics. The executable guard checks query/source counts,
k, self-exclusion and domain identity semantics, coordinate and field
shape/dtype/device/stride, field names, and the absence of gradient-bearing
inputs. Coordinates and feature values remain live operands, not cached graph
results. This separates changing data from changing programs and removes
repeated capture/lowering/provider preparation for compatible dynamic calls.

\subsection{Programs and bounded control}
Straight-line message-passing calls can be captured as a typed acyclic
\code{GraphProgram}. Independent legal leaves may fuse horizontally; dependencies
remain ordered. \code{@tg.jit} also captures supported Python loop forms, while
\code{repeat} and \code{while\_loop} offer explicit structured control over native
values. Loop-carried shape, dtype and device are invariant. Bounded while requires
a scalar boolean condition and a finite maximum iteration count. CPU lowering
supports the implemented structured loop slice; general multi-state CUDA loop
plans and structured reverse-loop lowering remain incomplete.
\FloatBarrier

\section{Differentiation and numerical contracts}
\label{sec:differentiation}
Forward execution computes outputs from input fields and parameters. Reverse-mode
automatic differentiation propagates an output cotangent back through that
computation~\cite{autodiff}. For a program with Jacobian $J_F$, the VJP is
$$
y=F(x),\qquad \bar{x}=J_F(x)^{\mathsf{T}}\bar{y}.
$$
The bar denotes a loss derivative: for a scalar loss, the output cotangent is
the derivative of that loss with respect to $y$. The full Jacobian need not be
materialized. Tiga implements this operation for supported native computations
and integrates differentiable Torch-facing calls with PyTorch autograd.

For fixed relation indices, weighted sum has a simple adjoint. With output
cotangent denoted by a bar, source and edge derivatives are
$$
 \bar{x}_j=\sum_{e:s(e)=j}w_e\bar{y}_{d(e)},\qquad
 \bar{w}_e=x_{s(e)}\bar{y}_{d(e)}.
$$
Repeated source indices require accumulation, not assignment. Empty destination
rows contribute no source/edge derivative. The example in Section~\ref{sec:programming-model} provides
an independently checkable forward and backward value contract.
For its summed-output loss, the output cotangents are all one. A source that
appears on multiple edges receives the sum of their weighted contributions;
each edge weight receives the feature value from its own source.

\subsection{Native and Torch paths}
Native reverse-mode transforms construct derivative expressions from supported
Tensor and relation operations. Torch-facing calls preserve ordinary Torch
outputs and autograd integration, but their backward implementation is path-specific.
Some compiled CSR forwards replay fixed-topology Torch semantics during backward.
That route is differentiable without being a compiled TTIR backward kernel.
Explicitly wrapping a Torch tensor into a native tensor shares storage but does
not merge the two autograd histories.

Supported traced edge neural networks have a CUDA tile forward and a symbolic-VJP
recompute kernel. The tests compare output and gradients of source features,
positions and every module parameter to an eager reference. CPU tests additionally exercise explicit edge
attributes with double-precision finite differences. These checks establish the
tested first-order envelope, not arbitrary Python capture or universal
higher-order differentiation. ReLU-like nondifferentiabilities need appropriate
test points, rather than unqualified finite-difference equality at a kink.

\begin{table}[htbp]
\centering\small
\begin{tabularx}{\linewidth}{>{\raggedright\arraybackslash}p{0.24\linewidth} Y Y}
\toprule
Execution path & Forward & Backward \\
\midrule
Torch CSR & Compiled or adapter execution & Torch autograd; some routes replay fixed-topology Torch operations \\
CUDA EdgeNN & Fused edge evaluation and reduction & Symbolic VJP with activation recomputation; input, position and parameter gradients \\
CUDA streaming attention & Implicit dense or triangular relation; tiled online softmax & No streaming backward; explicit Torch reference supports semantic gradient checks \\
Native CPU CSR and paging & Supported native operations and paged reducers & Native VJP, including supported disk-backed fields \\
CUDA paged CSR & FP32 paged execution with resident output & Not implemented on this path \\
Generated exact kNN & Fused selection and aggregation & No generated backward on this optimized path \\
Distributed native programs & Local computation after halo exchange & Local VJP plus reverse halo accumulation on supported CPU/CUDA paths \\
\bottomrule
\end{tabularx}
\caption{Differentiation follows the selected execution path, not just the graph
constructor. These are capability boundaries; the primary performance
experiments measure forward execution, while Section~\ref{sec:supplement}
adds a differentiated EdgeNN workload.}
\label{tab:gradients}
\end{table}

For application code, set \code{requires\_grad} before evaluation and call
\code{torch.autograd.grad} or \code{loss.backward()} on ordinary Torch outputs.
Native tensors use \code{tg.autograd.grad}. Gradients are taken with respect to
values on the selected topology: radius cutoff membership and kNN index
selection are discrete and are not themselves differentiated. The EdgeNN
example in Appendix~\ref{app:edge-nn} checks feature, position and parameter
gradients against an explicit reference. The billion-edge capacity measurement
uses CUDA paging and therefore establishes forward capacity, not billion-edge
training capacity.

\subsection{Online softmax and approximation}
For messages consisting of a score and value, the streaming state tracks the
maximum score, normalized mass and weighted value accumulator. Two nonempty
states combine by rescaling into a common maximum:
$$
 m=\max(m_a,m_b),\qquad
 \ell=e^{m_a-m}\ell_a+e^{m_b-m}\ell_b,\qquad
 o=e^{m_a-m}o_a+e^{m_b-m}o_b.
$$
Finalization divides the accumulator by the mass for nonempty rows; empty-state
handling must avoid undefined subtraction of infinities and follow the
implementation's reducer convention. This numerical structure is related to
IO-aware attention algorithms~\cite{flashattention}; it does not imply that every
Tiga path achieves their performance or full feature coverage.
Appendix~\ref{app:attention} gives an executable causal-attention example and
separately checks its compiled forward and Torch semantic gradients.

An optional positive pruning threshold enables an approximate dense-attention
forward path. Scores and key reads are still needed for the pruning decision;
eligible tiles skip value loading and accumulation. A large score gap can make
a tile's contribution small, but a block-level criterion is not a per-query
error bound. Threshold choice affects accuracy and work saved, and may not improve
latency. The exact default is a disabled threshold, not a threshold of zero.
Pruned backward is not claimed. Exact and approximate workloads must not be mixed
in a single speedup comparison without reporting numerical error.

\section{Storage, memory and distribution}
\label{sec:runtime}
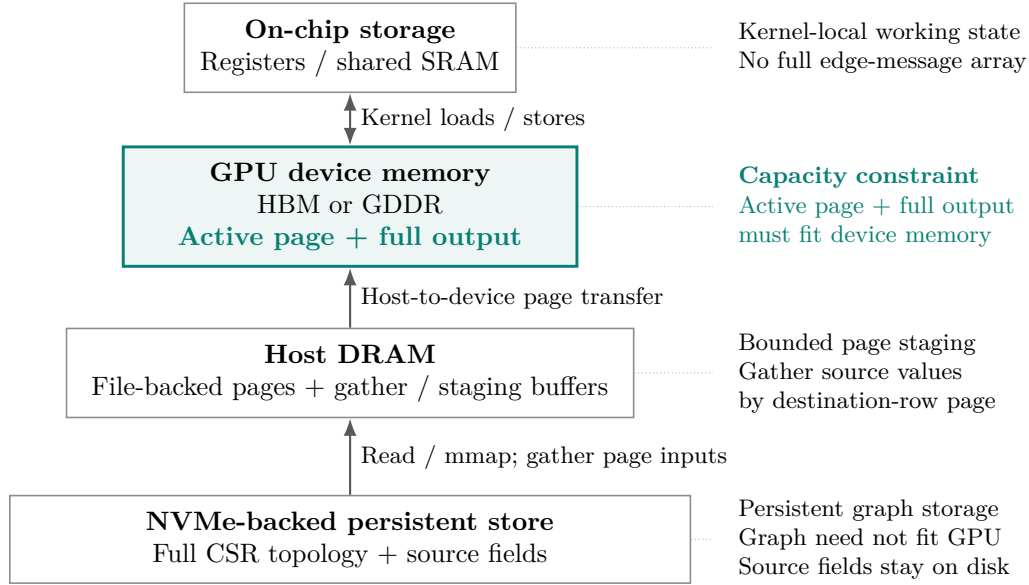
\begin{figure}[htbp]
\centering
\begin{tikzpicture}[font=\small, >=Latex,
  tier/.style={draw=black!45, line width=.6pt, align=center, inner sep=6pt},
  note/.style={anchor=west, align=left, text width=4.3cm, font=\footnotesize},
  transfer/.style={->, line width=.8pt, draw=black!65}]
  \node[tier, minimum width=3.6cm, minimum height=1.0cm] (chip) at (4.5,6.5)
    {\textbf{On-chip storage}\\Registers / shared SRAM};
  \node[tier, draw=tigateal, fill=tigateal!7, line width=1.1pt,
    minimum width=6cm, minimum height=1.25cm] (gpu) at (4.5,4.4)
    {\textbf{GPU device memory}\\HBM or GDDR\\
     \textcolor{tigateal}{\textbf{Active page + full output}}};
  \node[tier, minimum width=7.5cm, minimum height=1.1cm] (ram) at (4.5,2.2)
    {\textbf{Host DRAM}\\File-backed pages + gather / staging buffers};
  \node[tier, minimum width=9cm, minimum height=1.1cm] (disk) at (4.5,0)
    {\textbf{NVMe-backed persistent store}\\Full CSR topology + source fields};

  \draw[transfer] (disk.north) -- node[right, font=\footnotesize]
    {Read / mmap; gather page inputs} (ram.south);
  \draw[transfer] (ram.north) -- node[right, font=\footnotesize]
    {Host-to-device page transfer} (gpu.south);
  \draw[<->, line width=.8pt, draw=black!65] (gpu.north) --
    node[right, font=\footnotesize] {Kernel loads / stores} (chip.south);

  \node[note] at (9.5,6.5) {Kernel-local working state\\No full edge-message array};
  \node[note, text=tigateal] at (9.5,4.4)
    {\textbf{Capacity constraint}\\Active page + full output\\must fit device memory};
  \node[note] at (9.5,2.2) {Bounded page staging\\Gather source values\\by destination-row page};
  \node[note] at (9.5,0) {Persistent graph storage\\Graph need not fit GPU\\Source fields stay on disk};
  \draw[black!30, densely dotted] (chip.east) -- (9.3,6.5);
  \draw[black!30, densely dotted] (gpu.east) -- (9.3,4.4);
  \draw[black!30, densely dotted] (ram.east) -- (9.3,2.2);
  \draw[black!30, densely dotted] (disk.east) -- (9.3,0);
\end{tikzpicture}
\caption{Storage hierarchy for paged graph execution. The central
constraint is \emph{active page plus full output}, not full-graph residency on
the GPU. Arrows show input movement and kernel access; the implementation also
returns each completed output page through host bytes before writing it into
the resident device output. No direct NVMe-to-GPU path is implied. Widths are
schematic, not hardware capacities or bandwidths. Experimental hardware and
measured footprints are reported in Section~\ref{sec:experimental-setup} and
Section~\ref{sec:billion}, respectively.}
\label{fig:storage-hierarchy}
\end{figure}

The storage hierarchy makes the capacity strategy explicit
(Figure~\ref{fig:storage-hierarchy}): keep the complete relation and source
fields in a persistent store, stage one destination-row page through host
memory, and execute with a bounded temporary GPU working set. The final output
remains in device memory. FlashAttention's IO-aware view distinguishes on-chip
storage from device memory~\cite{flashattention}; here the paging boundary also
crosses host DRAM and NVMe. Tiga's paging mechanism bounds graph residency across
these outer memory levels; tile-local fusion bounds intermediate storage within a kernel.

\subsection{Logical values versus physical residence}
\label{sec:gpu-paging}
The native runtime records value metadata and physical buffers separately.
Execution scopes select default device, tier budgets, spill directory and graph
paging defaults; they do not migrate preexisting values. Memory reports count
scope-created allocations, not process RSS or Torch-owned storage. Limits are
budgets, not reservations of physical capacity.

Ordinary Torch storage remains under Torch's allocator. Applications that need
Tiga-managed paging or distributed ownership use native \code{tg.Tensor}
fields for those paths, keeping the same message/reducer definition. The
\href{https://walkerchi.github.io/TIGA-lang/memory/}{memory API} and
\href{https://walkerchi.github.io/TIGA-lang/examples/distributed-memory/}{paged and distributed examples}
show the allocation scopes, graph snapshots and rank setup used by this runtime.

Temporary spill releases the tensor's resident-buffer reference while retaining
shape, dtype, logical device, version and same-process gradient history.
Observation restores the value. Other views or launches may still own storage,
so logical eviction is not proof that all associated device bytes were freed.
Persistent snapshots store values, not a reconstructible autograd graph. Device
copy is differentiable on supported native paths but currently eager, not merely
a scheduler hint. The optional LRU policy evicts whole idle native tensors; it
does not tile a working set larger than device memory, manage Torch allocations
or supply transparent GPU-to-NVMe execution for arbitrary kernels.

Paged CSR streams bounded destination-row ranges on CPU and, for the tested
FP32 forward slice, on CUDA via host staging. CPU paging has a VJP path;
CUDA paged backward remains unsupported. Graph offload and
field storage have distinct memory lifetimes; automatically paging topology
does not imply every node/edge field is paged. Forward output assembly allocates
the final resident tensor once and writes each completed page into its row
range. Only one output-page payload is retained on the host; the final output
must still fit the selected device. Persistent CSR ingestion in the large-graph
experiment writes bounded chunks directly to the versioned store, without
constructing a graph-sized Python index list.

This bounded-memory path is currently a capacity mechanism rather than an
optimal IO pipeline. Source fields are gathered on the host into page-edge
order and copied to CUDA, so repeated source IDs can duplicate feature traffic
within a page. Each completed output page is copied back to host bytes before
being written into the final device output. Page-local staging bounds the
temporary working set but does not eliminate these copies. Section~\ref{sec:billion} measures their
cost and separates it from GPU kernel execution. Deduplicating source staging,
direct device output assembly and asynchronous IO are distinct optimizations;
they are not implied by an offload interface.

The optional runtime prefetcher maintains a bounded queue of topology-page
reads and contiguous-field read-ahead hints. It does not move source gathers,
host packing or CUDA copies into a compiler-scheduled asynchronous pipeline.
Consequently prefetch depth is not a guarantee of disk/PCIe/kernel overlap;
the capacity experiment reports the explicit serialized configuration.

\subsection{Ownership, halos and progress}
A rank is a cooperating worker process. A logical device mesh and destination
partition assign output ownership to ranks. Remote source values needed by local
destinations are held as ghost copies; the halo map specifies which values must
cross rank boundaries. The public schedule completes halo exchange before computing all owned
destination rows in one local call, without split-output assembly.
Reverse-mode execution accumulates ghost contributions back
to owners. Event ordering is a correctness requirement independently of whether
communication and computation actually overlap in time.
For example, if a destination on rank 0 reads a source owned by rank 1, the
forward exchange copies that source value to rank 0. Backward sends the source's
gradient contribution in the opposite direction and adds it to the gradient on
rank 1. This is why ownership must be retained through differentiation, not
treated solely as a forward communication detail.

The runtime supports CPU rank-local execution, MPI transport, and two-host CUDA
forward/VJP through TCP and NCCL. Deployment supplies worker processes, rank
membership and a communicator. Ownership remains fixed during execution;
Section~\ref{sec:two-host} evaluates an equal-destination partition on spatial meshes.

\subsection{Validation and trust boundaries}
Dispatch distinguishes supported compiled paths from reference/provider routes.
Full CSR validation checks monotonic row pointers, endpoints and source-index
bounds; basic validation checks metadata and is not an exhaustive scan.
Paged reads validate each consumed range. Tensor snapshots, geometry importers
and TCP transports are not security sandboxes: compiled UDFs and providers
execute trusted native code, and transport deployment requires trusted peers.
\FloatBarrier

\section{Evaluation: speed, capacity, and distributed execution}
\label{sec:evaluation}
The evaluation measures runtime performance and allocated memory against matched
PyTorch~\cite{pytorch} and PyTorch Geometric (PyG)~\cite{pyg,pyg2} implementations,
single-GPU capacity with storage offload, and the
latency of distributed graph execution.
Appendix~\ref{app:edge-nn} provides a differentiable EdgeNN example and a
Torch forward/gradient reference. Numerical gradient validation is distinct from
the forward-performance measurements reported here.

\subsection{Setup and comparison boundaries}
\label{sec:experimental-setup}
The single-GPU host has a Ryzen 7 255, approximately 44 GiB RAM, local NVMe
storage and an RTX 5070 Ti with 16 GiB device memory (driver 595.84).
The Torch path uses Torch 2.11.0+cu128, Triton 3.6.0, PyG 2.8.0.post1 and
pyg-lib 0.9.0+pt211cu128. The second host supplies an RTX 4070 Ti SUPER with
16 GiB under Ubuntu/WSL. Existing services remain running, occupying roughly
3 GiB locally and 12 GiB remotely. These are shared hosts, not dedicated,
clock-locked machines. GPU memory here means device memory; these consumer
GPUs use GDDR. Distributed communication uses NCCL 2.28.9 with its Socket
transport across the two hosts. The reproduction archive specifies the network configuration.

Inputs, dimensions, precision, outputs and graph semantics are matched within
each comparison. The primary comparisons are FP32 forward evaluations;
Section~\ref{sec:supplement} additionally measures differentiated EdgeNN steps.
Performance comparisons use ordinary Torch tensors and public Tiga
calls; paging and distributed experiments use native tensors. Timers cover synchronized host wall
time through completed output, with numerical validation outside the timer.
First compilation, warm execution and profiling have separate scopes. Speedup is
$$
 S = \frac{T_{\mathrm{matched\ baseline}}}{T_{\mathrm{Tiga}}}.
$$
Values below one indicate slowdown. Source/compiler hashes, commands, versions
and raw samples accompany each experiment's source snapshot.

\subsection{Runtime performance and memory efficiency}
\label{sec:performance}
Three workloads isolate different uses of the graph abstraction:
\begin{itemize}
\item Stored CSR: degree 32, 32 features, weighted neighbor sum; 8,192,
32,768 and 131,072 nodes. Peers are \code{torch.sparse.mm} and PyG
\code{MessagePassing} with COO gather--scatter. CSR setup is untimed; PyG's
required COO representation is included in its memory. The sparse peer prevents
one COO path from standing in for all optimized PyG execution.
\item Dynamic radius: scalar distance-weighted sum, nominal degree 32;
8,192, 32,768 and 131,072 points. Every call copies one of two different coordinate
snapshots, rebuilds the relation and consumes it. Peers are PyG's radius builder
plus COO message passing and pure Torch chunked-distance CSR construction plus
Torch sparse multiplication. The generated path avoids accepted-edge, distance and
edge-message arrays.
\item Dynamic exact kNN: scalar neighbor sum, $k=16$; 1,024, 4,096 and 16,384
points, extended to 32,768, 65,536 and 131,072 points. The practical peer is PyG's
kNN builder plus COO message passing. The pure Torch peer computes direct squared
distances in row chunks, selects top-k, then gathers and sums features.
Neither Torch baseline calls a Tiga graph builder or uses approximate search.
\end{itemize}

Both dynamic snapshots undergo complete edge-set comparison between the explicit
reference and PyG before timing. Radius coordinates use a binary grid and a
cutoff between squared-distance levels to avoid ambiguous FP32 boundaries.
The radius neighbor cap is 128; the complete comparison establishes that it
truncates neither snapshot. kNN reference distances use direct differences,
not the matrix-product identity. Every timed output is checked at relative and
absolute tolerance $3\cdot10^{-4}$. Coordinates actually change; unchanged-input
cache lookup is not counted as rebuilding. These generated dynamic paths use
scalar features, not wide-feature edge networks.

The two Torch graph builders bound each distance block to 16,777,216 pairs
(64 MiB for one FP32 distance plane; multiple temporary planes may coexist).
They still inspect all candidate pairs and materialize the complete accepted
relation. At 131,072 points an unchunked distance plane alone would require
64 GiB; chunking permits a real measured baseline instead of a predicted OOM.
This is an eager all-pairs implementation, not an optimized spatial index or a
claim that Torch cannot express a better algorithm.

Each of 36 configurations runs in a fresh process, with provider order randomized
across the matrix, four warmups and ten checked timed calls. Curves show median
and sample range, not cross-process confidence intervals. Peak Torch-allocated
GPU bytes include live inputs, coordinate fixtures, output, provider-specific
representations and temporary tensors. They exclude allocator reserve, CUDA
context/module storage and other processes. Counters are reset per measured
call after warmup; reserved-memory peaks are not substituted for allocation.
Separate largest-shape audits found zero additional Tiga-native device-buffer
allocations on each of the three Torch paths.

\begin{figure}[htbp]
\centering
\includegraphics[width=\linewidth]{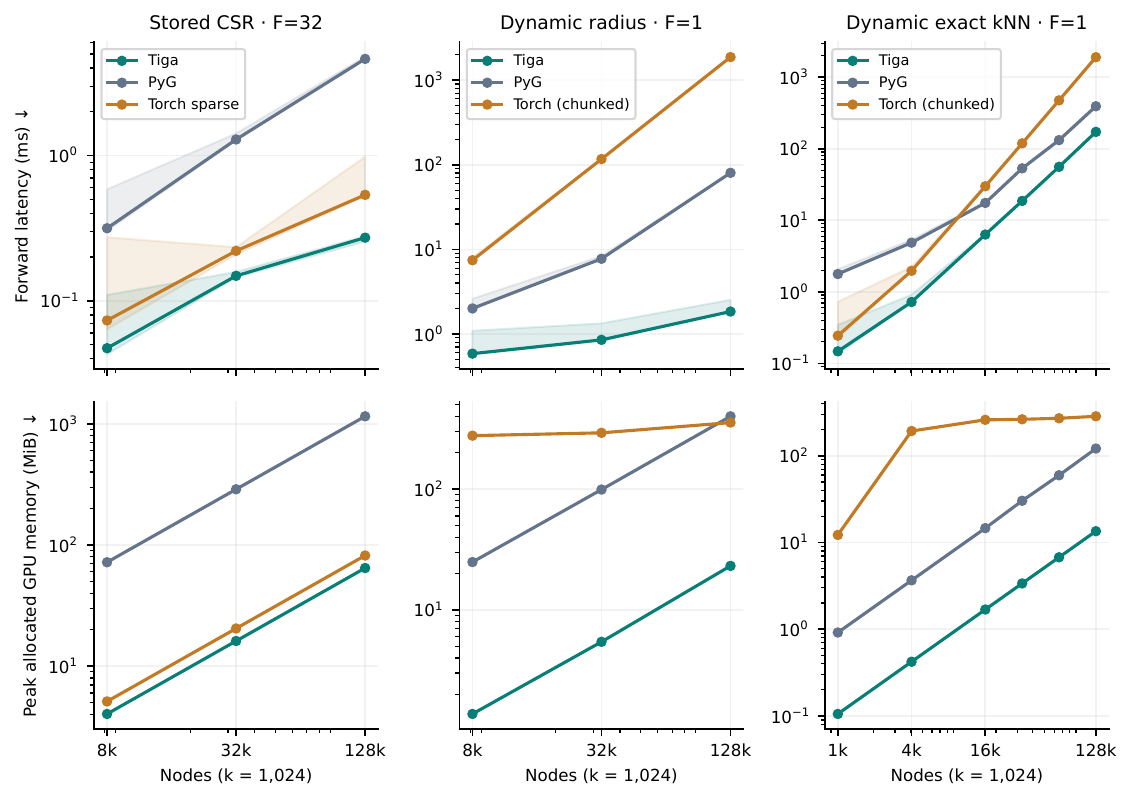}
\caption{Matched forward latency and measured allocated GPU memory.
Columns are different workloads; feature widths and sizes are not pooled.
Both rows use logarithmic axes and lower is better. Dynamic calls rebuild and
consume changed coordinates. Torch (chunked) means a pure Torch graph builder
and Torch aggregation, not a Tiga-built graph. kNN extends through 128K points;
the selected graph at that size has approximately 2.10M directed edges.}
\label{fig:q1}
\end{figure}

Stored CSR takes approximately 0.047--0.272 ms, giving 1.49--1.97$\times$
speedup over Torch sparse on the three measured sizes. Allocation is about
1.27$\times$ lower than the sparse peer, rather than the much larger reduction
against materialized COO messages. The largest dynamic radius case takes
1.84 ms versus PyG's 80.24 ms and pure Torch's 1,870.09 ms;
allocation peaks are 23.14, 400.22 and 355.01 MiB respectively.
These are workload-specific outcomes, not a framework ranking.

Exact kNN uses semantic executable rebinding and the exact tile-selection
pruning described in the compiler section. At 1,024 points Tiga takes 0.147 ms
versus 0.244 ms for pure Torch and 1.775 ms for PyG. At 16,384 points the
times are 6.32, 29.92 and 17.50 ms, respectively. At 131,072 points they are
172.37, 1,908.43 and 393.21 ms: Tiga is 2.28$\times$ faster than the PyG peer
with 13.50 versus 121.13 MiB allocated (pure Torch: 285.00 MiB).
Across the six measured sizes the PyG-relative speedup is 2.28--12.06$\times$.
All paths rebuild relations from changed coordinates; executable reuse does
not reuse selected neighbors. Candidate-distance work remains quadratic, so
these results do not establish subquadratic search or a universal advantage
over spatial-index kNN implementations.
\FloatBarrier

\subsection{Billion-edge execution on a single GPU}
\label{sec:billion}
The capacity workload is an explicit periodic directed CSR graph with 16
neighbors per destination and 32 FP32 features. Every edge is persisted and
traversed; the periodic closed-form oracle is used only for validation.
The sweep stops at 10M, 100M and 1B edges by design, not at a demonstrated
maximum supported size. Three fresh processes per size are retained from the
fully checked capacity run. Bounded ingest is untimed; forward includes
page loading, host staging, JIT as encountered, computation and output assembly,
but excludes full-output checking.

Pages contain 65,536 destination rows, prefetch is disabled, and the native
device budget is 12 GiB. This is a serialized capacity baseline, not a
measurement of overlapped IO or compiler-scheduled prefetch. The runtime's
optional read-ahead covers topology loading and contiguous field hints;
source gathering and GPU staging remain on the consuming thread. Enabling
that option alone does not create a disk--host--device double-buffered pipeline.
The final output remains resident. At 1B edges,
62.5M nodes require 8.5 GB of indices, 8 GB of source features and 8 GB of output.
The full-residency lower bound is 22.82 GiB with int64 CSR, or 18.86 GiB with
int32 CSR, both beyond 16 GiB. These are arithmetic infeasibility bounds, not
measured OOM attempts or Torch/PyG runs at this size. A hand-chunked peer could
also exchange storage traffic for capacity; paging is not claimed to be unique.

\begin{figure}[htbp]
\centering
\includegraphics[width=\linewidth]{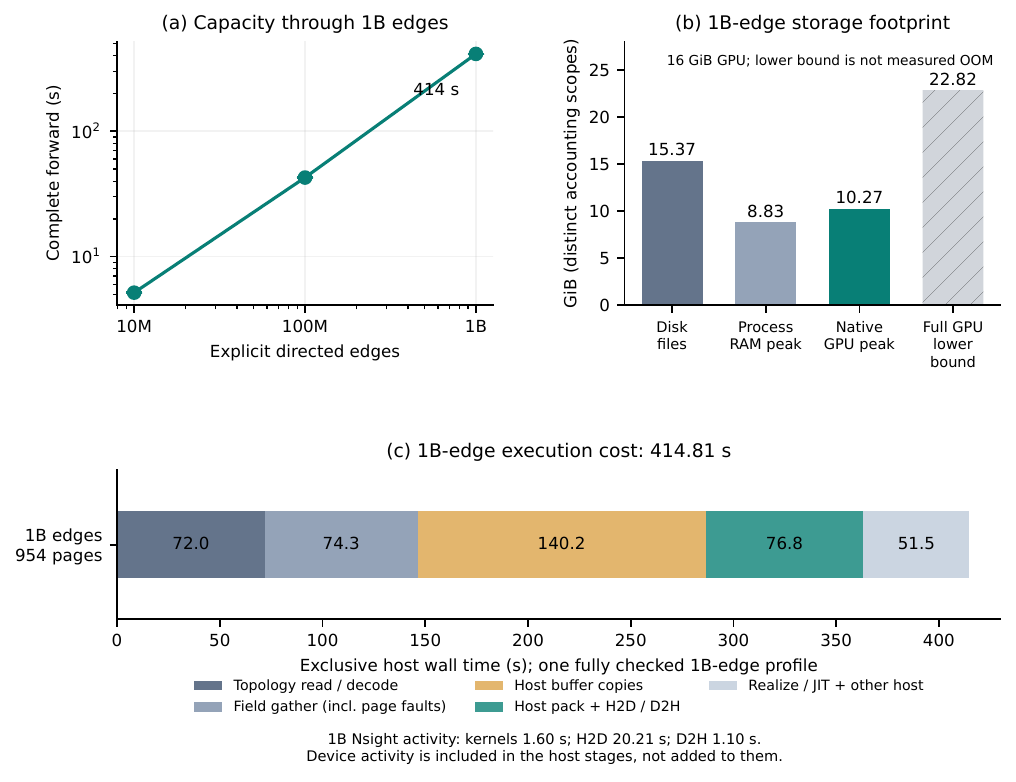}
\caption{Single-GPU capacity and execution cost. (a) Median/range of three complete calls per size.
(b) Disk footprint, process RAM peak, native GPU peak and calculated full-residency
requirement at 1B, with distinct accounting scopes. (c) Exclusive host stages in
one fully checked, instrumented 1B-edge forward call. Nsight
CUDA activity is already contained in the host intervals, not an extra stage.}
\label{fig:q2}
\end{figure}

The 1B median is 414.25 seconds (411.36--415.61), with maximum native GPU peak
10.27 GiB and forward process RSS 8.83 GiB. All two billion output scalars are
checked exactly. Topology and source occupy 15.37 GiB on disk. Device-wide
sampled usage, including services and allocation pools, reaches 13.25 GiB;
the native counter alone is not total physical usage. The output occupies
7.45 GiB, so capacity still depends on node count and feature width.

A separate instrumented 1B-edge call records exclusive nested host intervals
and an Nsight CUDA capture restricted to forward. All two billion output
scalars are checked exactly afterward. The profiled window takes 414.81 seconds;
the outer call takes 417.05 seconds including profiler start/stop overhead.
It is not pooled into the original three-trial latency sweep.
Topology read/decode takes 71.98 seconds, field gathering 74.27 seconds,
host-buffer copying 140.22 seconds, packing and host/device transfer stages
76.79 seconds, and realization/setup/assembly 51.55 seconds.
These exclusive host intervals sum to the profiled window. Nsight separately
records 1.60 seconds of CUDA kernels, 20.21 seconds H2D and 1.10 seconds D2H;
these device activities are already contained in the host stages.

Host staging, rather than kernel arithmetic, dominates. Source staging expands
feature rows at edges: an 8 GB stored source field generates 128 GB of gathered
values. Total H2D traffic is 144.50 GB; D2H is 8 GB. Eviction hints on only the
three fixture files yield 16.50 GB of process physical reads, including any
runtime/compiler file reads. mmap gathering includes page-fault IO, not an
isolated disk-service timer. The graph fits host RAM; this is not evidence for
a graph beyond RAM capacity. CUDA paged backward, 1B recurrence and output
offload remain outside the demonstrated result.

\paragraph{Page-size sensitivity.}
A supplementary 10M-edge sweep holds the graph, 32-feature operation and 12 GiB
budget constant while varying destination rows per page. Nine fresh processes
(three per setting, randomized order) run the same public paged path with
prefetch disabled. Each checks all 20 million output scalars exactly.
Forward includes JIT as encountered; ingestion and validation are excluded.
The OS page cache is not flushed, so this is not a cold-NVMe measurement.
\begin{table}[htbp]
\centering\small
\begin{tabular}{rrrr}
\toprule
Rows per page & Pages & Forward (s) & Native peak (MiB) \\
\midrule
16,384 & 39 & 4.95 (4.94--4.96) & 435.5 \\
65,536 & 10 & 4.87 (4.81--4.90) & 935.3 \\
262,144 & 3 & 5.41 (5.35--5.46) & 1442.1 \\
\bottomrule
\end{tabular}

\caption{Page-size sensitivity at 10M edges. Forward shows median and range of
three complete calls; native peak is the maximum scoped device allocation.
It excludes CUDA context, external services and OS page cache.}
\label{tab:page-sensitivity}
\end{table}
Larger pages increase peak allocation without monotonically improving time:
65,536 rows are slightly faster here than either alternative. Fixed setup and
host-staging costs limit gains from reducing page count. This small-graph study
does not identify an optimal 1B schedule, establish capacity beyond host RAM,
or compare against another bounded-memory implementation.
\FloatBarrier

\subsection{Distributed execution and communication overhead}
\label{sec:two-host}
The 5070 Ti and 4070 Ti SUPER each own half the destination rows of one graph,
exchanging remote source features through halo maps. This is two-rank graph
partitioning, not feature-axis tensor parallelism or independent data-parallel
graphs. Ownership is fixed for each run.

The workload is a nonperiodic structured three-dimensional hexahedral mesh.
Nodes sharing an element are connected in both directions, excluding self
entries: an interior node has 26 neighbors, and exterior nodes have fewer.
This matches the connectivity of a trilinear hexahedral FEM discretization; the measured
operation is 16-feature neighbor summation, not stiffness assembly or a complete
FEM solve. Cubes have side lengths 32, 64 and 96, giving 32,768, 262,144 and
884,736 nodes and 797,816, 6,596,856 and 22,508,920 directed edges.

Lexicographic x-major numbering makes the equal-row partition a spatial plane.
Each rank owns half the volume and exchanges one interface layer rather than
a volume-sized ghost set. For a cube of side length L,
$$
N=L^3,\qquad N_{\mathrm{halo,rank}}=L^2,\qquad
f_{\mathrm{boundary}}=\frac{2}{L}.
$$
The boundary
fractions decrease from 6.25\% to 3.125\% and 2.083\%; total traffic across both
ranks is 0.125, 0.500 and 1.125 MiB per call. Thus the experiment measures the
small surface-to-volume regime relevant to spatial graphs.

Each GPU also executes the complete graph alone. The \emph{communication then
compute} schedule completes halo exchange before computing local outputs.
Two warmups precede five changed-input, fully checked calls. Per-sample distributed latency
is the maximum of paired rank durations, followed by a median, not the mean of
rank medians. Topology setup, input allocation, the control barrier and checking
are excluded. Setup topology is replicated and partition maps cached: this is
strong scaling, not distributed topology-capacity evidence beyond one GPU.

Both hosts use matching Tiga sources and compiler executables.
Single-card and two-card modes share the native-input timing boundary; their
times are not directly comparable with the different ordinary-Torch workloads
in Section~\ref{sec:performance}.
Modes run in fixed order; small sample counts and shared services limit
cross-run generalization.

\begin{figure}[htbp]
\centering
\includegraphics[width=\linewidth]{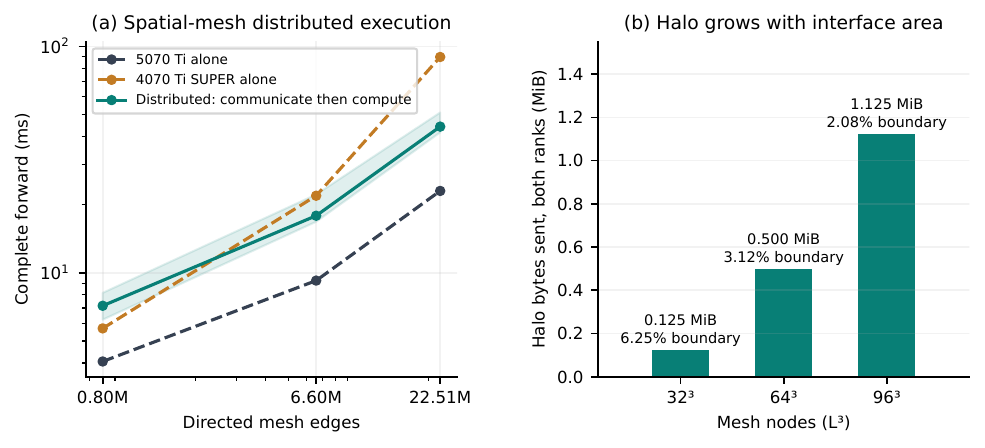}
\caption{Distributed NCCL execution of spatial mesh graphs on two hosts.
(a) Full-graph single-GPU baselines and paired two-rank latency.
(b) Halo traffic grows with the interface area while the boundary fraction
shrinks with mesh size. Latency covers the complete forward call: halo exchange,
runtime scheduling, computation and synchronization. Each rank owns half the destination nodes.}
\label{fig:q3}
\end{figure}
At 22.51M directed edges, the 5070 Ti and 4070 Ti SUPER single-GPU medians are 23.00 and 89.75 ms. The communication-then-compute schedule takes 44.22 ms on the paired-rank critical path, 1.92 times the faster single-GPU latency. Only 2.08\% of each rank's destination rows touch the interface, and total halo traffic is 1.125 MiB. This fixed partition assigns half the destination rows to each GPU. Compute-aware partitioning is a potential optimization for the measured heterogeneous pair.

\FloatBarrier

\subsection{Controlled execution studies}
\label{sec:supplement}
The following studies distinguish stored-topology execution, differentiated
message computation and first-call cost from the dynamic-search comparisons.
They use the same single-GPU host, with three fresh processes per configuration,
seeds 20260921--20260923, four warmups and ten checked calls per process.
Configuration order is randomized. Plots show the median and range of process
medians, not a confidence interval. Feature values alternate between two
snapshots; the input copy and checking are outside the synchronized wall timer.
Reported memory is peak Torch-allocated storage, including live topology,
inputs, outputs, tapes and workspaces, but excluding allocator reserve, CUDA
context and other processes. The source snapshot is independent of the older
experiments; unchanged hardware does not imply an unchanged implementation.

\paragraph{Holding topology constant.}
A common radius search produces a fixed CSR snapshot at 8,192 or 32,768
uniform binary-grid points, with nominal degree 32. The three consumers receive
identical indices, Euclidean weights and 32- or 128-component features: Tiga's
compiled weighted sum, Torch gather--multiply--scatter, and Torch sparse matrix
multiplication. Search is outside timing for all three. CSR hashes are checked
between peers in each trial. This is a topology-controlled execution-path
comparison, not a single-pass on/off ablation: traversal, dispatch and reduction
strategy can still differ.

\begin{figure}[htbp]
\centering
\includegraphics[width=\linewidth]{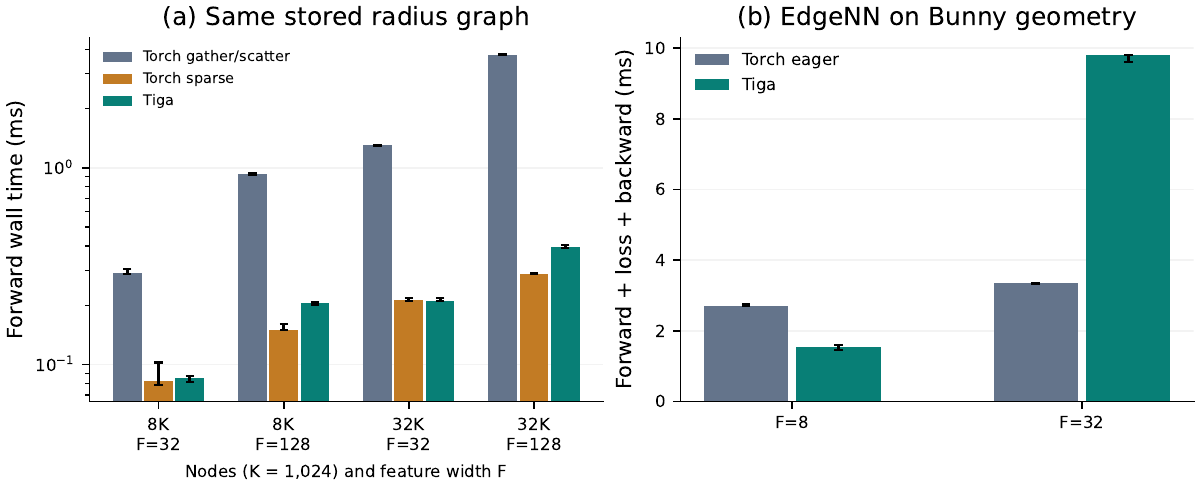}
\caption{Supplementary execution studies. (a) The same stored radius topology
is consumed by three implementations; graph construction is excluded.
(b) Differentiated EdgeNN steps on a radius graph of Stanford Bunny geometry.
Bars show medians of three process medians, with their ranges. Lower is better;
only panel (a) uses a logarithmic axis.}
\label{fig:supplement}
\end{figure}

At 32,768 nodes and 128 features, measured forward times are 0.395 ms, 3.759 ms, 0.289 ms for Tiga, Torch gather/scatter and Torch sparse, respectively. Median process-peak allocations are 67.3 MiB, 1007.3 MiB, 67.6 MiB. These measurements hold search and topology fixed; execution strategy and launch overhead still differ.

The reduction in intermediate storage is clear against explicit message
materialization; the optimized sparse-library peer can be faster than Tiga
and has similar storage. Thus eliminating edge messages is useful but does
not alone establish a better schedule than a sparse library. This study does
not isolate the separate gain from avoiding accepted-edge storage in the
generated-radius experiment.

\paragraph{Differentiation on real geometry.}
The Stanford Bunny zippered reconstruction provides 35,947 reconstructed
vertices~\cite{stanfordbunny}. Coordinates are centered and divided by the largest
bounding-box extent; radius 0.015 gives 356,260 directed non-self edges. The
reconstructed surface is the data source, not raw scanner noise. Features are
synthetic; an untrained MLP maps displacement and source features through a
32-unit ReLU layer to three outputs, summed at destinations. The mean squared
difference from normalized positions supplies a scalar regression loss.
Both paths use the same fixed graph and parameters. This workload measures a
differentiated geometric operator, not downstream task accuracy, convergence
or a complete optimizer step.

Every retained call checks outputs and gradients of features, positions and
all four parameter tensors against explicit Torch gather--MLP--scatter.
Position gradients hold edge membership fixed. In addition to elementwise
tolerances, each gradient must have relative L2 error below 0.002. The Tiga
path must report its compiled tile VJP; a reference fallback is rejected.
Table~\ref{tab:supplement-edgenn} separates forward from loss-plus-backward.
Independent medians of these stages need not sum exactly to the median total.

\begin{table}[htbp]
\centering\small
\begin{tabular}{llrrrr}
\toprule
Input width & Path & Forward (ms) & Backward (ms) & Total (ms) & Peak (MiB) \\
\midrule
8 & Torch eager & 0.64 & 2.07 & 2.71 & 172.25 \\
8 & Tiga & 0.86 & 0.72 & 1.54 & 32.34 \\
32 & Torch eager & 0.94 & 2.42 & 3.36 & 209.22 \\
32 & Tiga & 0.92 & 8.86 & 9.79 & 37.58 \\
\bottomrule
\end{tabular}

\caption{Bunny EdgeNN: medians across three process medians. Peak allocation is
the median of per-process measured peaks. Backward includes loss construction;
no optimizer update is timed.}
\label{tab:supplement-edgenn}
\end{table}

At width eight, the measured total improves from 2.71 to 1.54 ms with substantially
less allocated storage. At width 32, Tiga uses less storage but its 8.86 ms
backward makes the total slower than Torch. Recompute-based fusion is therefore
a memory--time tradeoff, not a universal training acceleration. This experiment
does not independently isolate recomputation, atomic accumulation and scheduling
as causes of that slowdown.

\paragraph{First result and amortization.}
For the 32,768-node, 32-feature stored-radius case, three additional worker pairs
use empty isolated Triton, TorchInductor and CUDA caches, then reuse each pair's
disk caches in a new process. Imports, CUDA initialization, graph setup and the
eager correctness oracle precede timing. The first result includes Tiga capture,
lowering, provider work, module loading and execution, rather than pure compile
time. A disk-cache hit does not restore an in-process executable.
The gather/scatter peer has already served as the correctness oracle, so its
entry is the first timed call, not cold library initialization. All entries
exclude that common setup; this is an execution-ready comparison rather than
time from launching Python to a first answer.

\begin{table}[htbp]
\centering\small
\begin{tabular}{lrr}
\toprule
Path & First result (ms) & Warm call (ms) \\
\midrule
Tiga, empty isolated caches & 1105.5 (1095.5--1113.8) & 0.214 \\
Tiga, reused disk caches & 635.1 (621.9--651.4) & 0.216 \\
Torch sparse, fresh process & 17.0 (16.9--17.5) & 0.213 \\
Torch gather/scatter, fresh process & 1.3 (1.2--1.3) & 1.301 \\
\bottomrule
\end{tabular}

\caption{First-result cost versus warm execution. First-result entries show
median and range of three fresh processes; warm entries are medians of process
medians. Common setup is excluded; the gather/scatter oracle is already warm.
Torch sparse uses its ordinary library environment.}
\label{tab:supplement-jit}
\end{table}

For first-result cost $C$, warm time $t$ and $K$ calls, an amortization model is
$$
T(K)=C+(K-1)t.
$$
The model assumes stable warm time; it is not a measured cumulative-time curve.
Here the cold first result costs about 1.11 s, and persistent-cache reuse still
costs about 0.64 s. Warm Tiga and Torch sparse are both about 0.21 ms, with no
measured steady-state advantage supporting an amortized crossover against the
sparse peer. Against explicit gather/scatter, warm savings can amortize setup
over repeated calls.
Using the reported medians, the model predicts a crossover against explicit Torch gather/scatter after approximately 1,018 calls. This is a model-derived estimate, not an observed crossover.

Short-lived programs should not be assigned warm-kernel
speedups without this qualification.
\FloatBarrier

\subsection{Reproduction and scope}
\path{data/comparison-v3/} stores the 36 comparison records, source snapshot
and allocation audits. \path{data/billion/} contains the nine capacity trials;
\path{data/paging-1b/} contains the instrumented billion-edge run and Nsight capture.
\path{data/distributed-spatial/} contains both rank records, commands, source
and interface-size checks for the spatial-mesh experiment.
\path{tools/build_three_questions.py} validates the archived configurations and
numerical checks before plotting. Rerunning hardware measurements uses the commands
and environments provided with each dataset.
\path{data/supplement/} adds the controlled execution and first-call records,
with \path{tools/run_supplement.py} and \path{tools/build_supplement.py} for
measurement and validation. The Stanford archive URL and checksum identify the
external geometry input; its model is not required to compile this report.
\path{data/page-sensitivity/} retains the nine page-size trials and commands;
\path{tools/build_page_sensitivity.py} validates and summarizes them.

The measurements support three conclusions: generated relations can reduce
runtime and intermediate storage; paging permits billion-edge execution beyond
device-memory capacity; and communication plus heterogeneous load balance
determines the cost of distributed execution. Gradient correctness is covered
separately by the differentiable examples and path-specific checks.
End-to-end model-training throughput, capacity beyond the measured edge count, and scaling to
larger device meshes remain subjects for further evaluation.

\clearpage
\section{Limitations and next steps}
The implementation has the following scope limits:
\begin{itemize}
\item Native and Torch paths have different operator, dtype and gradient coverage.
Traced modules support a guarded subset; structured reverse loops and general
higher-order differentiation remain incomplete.
\item Generated exact kNN supports a bounded FP32, small-k envelope. Larger-k
scratch, general metric lowering and generated backward require further work.
\item Implicit block composition, automatic heterogeneous out-of-core execution
and additional vendor backends remain incomplete.
\item Distributed execution uses fixed ownership. Compute-aware partitioning,
partition-local topology capacity, RCCL and failure recovery remain open.
\item Performance evaluation covers FP32 forward workloads and a fixed-topology
EdgeNN forward/backward step, not end-to-end model training. The latter shows a
memory reduction but a wide-feature backward slowdown. Sustained out-of-core training
and beyond-host-RAM graphs require additional measurement.
\end{itemize}

\section{Conclusion}
Tiga compiles graph message-passing programs while retaining the relations,
reducer algebra and derivative requirements that determine their execution.
This separation supports CPU and GPU lowering, forward and backward computation
on supported paths, and explicit control over storage residence and distributed
ownership. Generated relations reduce intermediate storage and improve runtime
on the measured workloads; paging enables billion-edge forward computation on
a memory-limited GPU. Distributed measurements show that a small halo alone is
insufficient for speedup in the measured heterogeneous deployment. Device speed,
host environment and competing services are not independently controlled, so
the observed gap does not isolate their individual effects.

The design links a local programming abstraction to the physical choices that
govern large-graph execution. Further work can build on the same representation
to extend backward coverage, reduce staging traffic and balance computation
across heterogeneous devices.

\section*{Reproducibility and tool use}
The report repository provides measurement records, numerical checks, source
snapshots, and figure-generation scripts. Each experiment identifies its timing
and allocation boundary; historical runs retain their original source snapshot.
Replotting recorded samples does not constitute a new hardware measurement.
The supplementary experiments include explicit compiler-path checks, and their
inputs and gradients are checked independently of the measurement window.

AI coding and writing assistants were used in software implementation and
testing, experiment design and tooling, literature discovery, interpretation
of measurements, figure/code preparation, and manuscript drafting and revision.
Numerical results originate from recorded program executions, not generated
estimates. Automated numerical and provenance checks were used to validate
outputs, supported gradients and the recorded evidence. The human author is
responsible for the final content, attribution and claims.

\clearpage
\appendix
\section{A differentiable EdgeNN example}
\label{app:edge-nn}
This example expresses a shared edge-local neural network followed by a
destination-wise sum. Positions and node features are ordinary
\code{torch.Tensor} objects; the weights are ordinary \code{torch.nn.Parameter}
objects. No native tensor wrapper or handwritten backward is required.
For a radius relation, the program computes
$$
\mathcal{N}(i)=\{j\ne i:\|p_j-p_i\|_2\le r\},\qquad
y_i=\sum_{j\in\mathcal{N}(i)}\left[
W_2\,\operatorname{ReLU}\!\left(W_1[p_j-p_i\,;\,x_j]+b_1\right)+b_2\right].
$$
Here the final bias is part of each edge message, inside the sum.
The feature widths are 8 for node inputs, 3 for displacement, 16 for the
hidden layer and 4 for each output. The reducer produces a zero vector for
a destination with no neighbors.

\subsection{Define the message and differentiate the result}
\code{tg.nn.trace} captures the supported Torch module without replacing its
parameters. Its multiple tensor arguments are concatenated along the feature
axis: \code{edge.displacement} supplies the first three features and
\code{src.x} the remaining eight. \code{reducer = tg.sum()} specifies aggregation
once on the class, rather than inside the edge function.
\begin{lstlisting}[language=Python,firstline=2,lastline=30]
import torch
from torch import nn
import tiga as tg

torch.manual_seed(7)
device = "cuda" if torch.cuda.is_available() else "cpu"
positions = torch.rand(128, 3, device=device, requires_grad=True)
x = torch.randn(128, 8, device=device, requires_grad=True)
mlp = nn.Sequential(nn.Linear(11, 16), nn.ReLU(), nn.Linear(16, 4)).to(device)


class EdgeNN(tg.MessagePassing):
    reducer = tg.sum()

    def __init__(self, module):
        super().__init__()
        self.mlp = tg.nn.trace(module)

    def edge(self, src, dst, edge):
        return self.mlp(edge.displacement, src.x)


graph = tg.Graph.radius(positions, cutoff=0.35)
program = EdgeNN(mlp)
y = program(graph=graph, src={"x": x}, dst={})
assert y.shape == (128, 4)
loss = y.square().mean()
variables = (x, positions, *mlp.parameters())
grads = torch.autograd.grad(loss, variables)
\end{lstlisting}

The call returns one four-component vector per destination. The loss yields
gradients for node features, positions and every MLP weight and bias through
\code{torch.autograd.grad}; \code{loss.backward()} is the usual alternative
when accumulated \code{.grad} fields are wanted. The position derivative acts
through the displacement of selected edges. Discrete radius membership is
not differentiated: moving across the cutoff changes the selected relation.

\subsection{Check against an explicit Torch reference}
The following continuation materializes edges only for validation, applies
the original Torch MLP, and aggregates with \code{index\_add}. Both forward
values and all gradients are compared on the same selected relation.
This isolates message, reducer and differentiation correctness; it is not an
independent validation of the radius-neighbor search.
\begin{lstlisting}[language=Python,firstline=32]
# Independent gather -> MLP -> index_add reference on the same relation.
row_ptr, col_idx = graph.resolve_csr()
rows = torch.repeat_interleave(
    torch.arange(128, device=device), row_ptr[1:] - row_ptr[:-1])
delta = positions[col_idx] - positions[rows]
messages = mlp(torch.cat((delta, x[col_idx]), dim=-1))
reference = torch.zeros_like(y).index_add(0, rows, messages)
reference_grads = torch.autograd.grad(reference.square().mean(), variables)
torch.testing.assert_close(y, reference, rtol=2e-4, atol=2e-5)
for actual, expected in zip(grads, reference_grads):
    torch.testing.assert_close(actual, expected, rtol=3e-4, atol=3e-5)
print(f"EdgeNN forward and all gradients checked on {device}.")
\end{lstlisting}

On the supported CUDA EdgeNN path, Tiga fuses edge evaluation and reduction
into tiled execution and recomputes local activations in the VJP, avoiding
a global edge-by-hidden-feature activation tensor. CPU execution remains
useful for functional checking but does not establish that CUDA performance
property. Unsupported module operations are outside this example's capture
contract. The complete runnable file is
\href{https://github.com/walkerchi/tiga-lang-paper/blob/main/examples/edge_nn.py}{\code{examples/edge\_nn.py}}; execute it
from the paper directory with Tiga and the appropriate Torch installation:
\begin{lstlisting}[language=TigaShell]
python examples/edge_nn.py
\end{lstlisting}

\clearpage
\section{Causal attention as message passing}
\label{app:attention}
Causal attention uses the same interface as sparse aggregation: the graph
selects admissible interactions, the edge function computes a score and value,
and the reducer normalizes their weighted sum. No attention-specific graph
operator is required. For each head, the computation is
$$
\mathcal{N}(i)=\{0,\ldots,i\},\qquad
s_{ij}=\frac{q_i^{\mathsf{T}}k_j}{\sqrt{D}},\qquad
y_i=\frac{\sum_{j\in\mathcal{N}(i)}e^{s_{ij}}v_j}
               {\sum_{j\in\mathcal{N}(i)}e^{s_{ij}}}.
$$
The destination supplies a query; each source supplies a key and value.
The triangular relation includes the diagonal: token zero reads only
itself; token one reads tokens zero and one. Causality therefore belongs to
the relation, not to an extra condition inside the edge function.

\subsection{Define and execute the program}
\code{tg.online\_softmax()} declares the reducer once. Calling
\code{self.reducer(score, value)} packages its two inputs; it does not run a
separate softmax for each edge. The compiler combines these messages over each
destination's neighbors using the stable state in
Section~\ref{sec:differentiation}. Pruning is disabled by default, so this
example evaluates the full causal relation, subject to floating-point rounding.
\begin{lstlisting}[language=Python,firstline=4,lastline=13]
import torch
import tiga as tg


class CausalAttention(tg.MessagePassing):
    reducer = tg.online_softmax()

    def edge(self, src, dst, edge, scale):
        score = (dst.query * src.key).sum(dim=-1) * scale
        return self.reducer(score, src.value)
\end{lstlisting}

Inputs and outputs are ordinary Torch tensors with logical shape
\code{(N, H, D)}: sequence length, number of heads and head width. The CUDA
streaming implementation currently expects head-major storage, obtained by
viewing a contiguous \code{(H, N, D)} allocation through \code{permute}.
This changes strides, not the public dimension order, and does not copy the data.
The runnable check defaults to \code{device="cuda"} and \code{nodes=65}, with
two heads of width 64. The sequence length exercises a partial tile.
\begin{lstlisting}[language=Python,firstline=27,lastline=38]
    torch.manual_seed(7)
    heads, width = 2, 64
    dtype = torch.float16 if device == "cuda" else torch.float32
    # Head-major storage, viewed as ordinary (N, H, D) Torch tensors.
    query, key, value = (
        torch.randn(heads, nodes, width, device=device, dtype=dtype)
        .permute(1, 0, 2) for _ in range(3))
    graph = tg.Graph.triangular(nodes, device=device)
    program = CausalAttention()
    with torch.no_grad():
        output = program(graph=graph, src={"key": key, "value": value},
                         dst={"query": query}, scale=width**-0.5)
\end{lstlisting}

On CUDA, the script checks the lowering reported by \code{program.explain()}.
It requires the compiled dense streaming path. This forward
traverses the implicit triangular relation without allocating a global
pairwise score or probability tensor. Changing the constructor to
\code{tg.Graph.dense(N)} expresses unmasked attention with the same edge
function and reducer.

\subsection{Validate outputs and query/key/value gradients}
An independent Torch implementation explicitly forms scores, masks future
positions, applies softmax and multiplies by values. This dense implementation
is a small correctness oracle, not the memory-efficient execution path.
\begin{lstlisting}[language=Python,firstline=16,lastline=23]
def torch_attention(query, key, value):
    # Public shape (N, H, D); matrix multiplication uses (H, N, D).
    q, k, v = (x.transpose(0, 1) for x in (query, key, value))
    scores = (q @ k.transpose(-1, -2)) * q.shape[-1]**-0.5
    future = torch.ones(q.shape[1], q.shape[1], device=q.device,
                        dtype=torch.bool).triu(1)
    weights = scores.masked_fill(future, -torch.inf).softmax(-1)
    return (weights @ v).transpose(0, 1)
\end{lstlisting}

The script compares the FP16 CUDA streaming output with this FP32 oracle.
The current streaming kernel is forward-only; setting
\code{requires\_grad} does not supply a compiled attention backward.
For a separate first-order derivative check, \code{program.reference}
explicitly executes the differentiable Torch semantics in FP32. Independent
leaf tensors and a random output cotangent test all three input gradients,
rather than just a summed-output loss.
\begin{lstlisting}[language=Python,firstline=47,lastline=62]
    variables = tuple(x.float().detach().requires_grad_()
                      for x in (query, key, value))
    q, k, v = variables
    semantic = program.reference(
        graph=graph, src={"key": k, "value": v},
        dst={"query": q}, scale=width**-0.5)
    reference_variables = tuple(x.detach().clone().requires_grad_()
                                for x in variables)
    reference = torch_attention(*reference_variables)
    cotangent = torch.randn_like(reference)
    gradients = torch.autograd.grad(semantic, variables, cotangent)
    expected_gradients = torch.autograd.grad(
        reference, reference_variables, cotangent)
    torch.testing.assert_close(semantic, reference, rtol=2e-4, atol=2e-5)
    for actual, expected in zip(gradients, expected_gradients):
        torch.testing.assert_close(actual, expected, rtol=3e-4, atol=3e-5)
\end{lstlisting}

The checks establish compiled-forward agreement and semantic gradient
agreement separately. The reference path materializes the selected edges and
their messages; it is intended here for small validation problems and does
not establish streaming training memory or throughput. A fused attention VJP
is a separate implementation requirement from the EdgeNN VJP in
Appendix~\ref{app:edge-nn}.

The complete file, including imports and assertions, is
\href{https://github.com/walkerchi/tiga-lang-paper/blob/main/examples/causal_attention.py}{\code{examples/causal\_attention.py}}.
With Tiga, its native compiler tools, Torch and the CUDA provider installed,
execute it from the paper directory:
\begin{lstlisting}[language=TigaShell]
python examples/causal_attention.py --device cuda
\end{lstlisting}
The optional \code{--device cpu} runs semantic checks only;
\code{--nodes} changes the sequence length. The CUDA check fails if the
streaming lowering is unavailable instead of accepting a reference fallback.

\clearpage
\bibliographystyle{plain}
\begingroup
\footnotesize\raggedright
\bibliography{references}

@inproceedings{actors,
  author = {Carl Hewitt and Peter Bishop and Richard Steiger},
  title = {A Universal Modular {ACTOR} Formalism for Artificial Intelligence},
  booktitle = {Proceedings of the Third International Joint Conference on Artificial Intelligence},
  year = {1973}, pages = {235--245},
  note = {\url{https://www.ijcai.org/Proceedings/73/Papers/027B.pdf}}
}

@inproceedings{pregel,
  author = {Grzegorz Malewicz and Matthew H. Austern and Aart J. C. Bik and James C. Dehnert and Ilan Horn and Naty Leiser and Grzegorz Czajkowski},
  title = {Pregel: A System for Large-Scale Graph Processing},
  booktitle = {Proceedings of the 2010 ACM SIGMOD International Conference on Management of Data},
  year = {2010}, pages = {135--146},
  note = {\url{https://research.google/pubs/pregel-a-system-for-large-scale-graph-processing/}}
}

@inproceedings{mlir,
  author = {Chris Lattner and Mehdi Amini and Uday Bondhugula and Albert Cohen and Andy Davis and Jacques Pienaar and River Riddle and Tatiana Shpeisman and Nicolas Vasilache and Oleksandr Zinenko},
  title = {{MLIR}: Scaling Compiler Infrastructure for Domain Specific Computation},
  booktitle = {2021 IEEE/ACM International Symposium on Code Generation and Optimization (CGO)},
  year = {2021}, pages = {2--14}, doi = {10.1109/CGO51591.2021.9370308},
  note = {\url{https://doi.org/10.1109/CGO51591.2021.9370308}}
}

@inproceedings{triton,
  author = {Philippe Tillet and H. T. Kung and David Cox},
  title = {Triton: An Intermediate Language and Compiler for Tiled Neural Network Computations},
  booktitle = {Proceedings of the 3rd ACM SIGPLAN International Workshop on Machine Learning and Programming Languages},
  year = {2019}, doi = {10.1145/3315508.3329973},
  note = {\url{https://www.eecs.harvard.edu/~htk/publication/2019-mapl-tillet-kung-cox.pdf}}
}

@inproceedings{pyg,
  author = {Matthias Fey and Jan Eric Lenssen},
  title = {Fast Graph Representation Learning with {PyTorch Geometric}},
  year = {2019},
  booktitle = {ICLR Workshop on Representation Learning on Graphs and Manifolds},
  note = {\url{https://arxiv.org/abs/1903.02428}}
}

@inproceedings{pyg2,
  author = {Matthias Fey and Jinu Sunil and Akihiro Nitta and Rishi Puri and Manan Shah and Bla{\v{z}} Stojanovi{\v{c}} and Ramona Bendias and Alexandria Barghi and Vid Kocijan and Zecheng Zhang and Xinwei He and Jan Eric Lenssen and Jure Leskovec},
  title = {{PyG} 2.0: Scalable Learning on Real World Graphs},
  booktitle = {Temporal Graph Learning Workshop at KDD}, year = {2025},
  note = {\url{https://arxiv.org/abs/2507.16991}}
}

@inproceedings{flashattention,
  author = {Tri Dao and Dan Fu and Stefano Ermon and Atri Rudra and Christopher R{\'e}},
  title = {{FlashAttention}: Fast and Memory-Efficient Exact Attention with {IO}-Awareness},
  year = {2022},
  booktitle = {Advances in Neural Information Processing Systems},
  volume = {35}, pages = {16344--16359}, doi = {10.52202/068431-1189},
  note = {\url{https://proceedings.neurips.cc/paper/2022/hash/67d57c32e20fd0a7a302cb81d36e40d5-Abstract-Conference.html}}
}

@inproceedings{graphiler,
  author = {Zhiqiang Xie and Minjie Wang and Zihao Ye and Zheng Zhang and Rui Fan},
  title = {Graphiler: Optimizing Graph Neural Networks with Message Passing Data Flow Graph},
  booktitle = {Proceedings of Machine Learning and Systems}, volume = {4}, year = {2022},
  note = {\url{https://proceedings.mlsys.org/paper_files/paper/2022/hash/a1126573153ad7e9f44ba80e99316482-Abstract.html}}
}

@inproceedings{featgraph,
  author = {Yuwei Hu and Zihao Ye and Minjie Wang and Jiali Yu and Da Zheng and Mu Li and Zheng Zhang and Zhiru Zhang and Yida Wang},
  title = {{FeatGraph}: A Flexible and Efficient Backend for Graph Neural Network Systems},
  year = {2020},
  booktitle = {SC20: International Conference for High Performance Computing, Networking, Storage and Analysis},
  note = {\url{https://arxiv.org/abs/2008.11359}}
}

@inproceedings{seastar,
  author = {Yidi Wu and Kaihao Ma and Zhenkun Cai and Tatiana Jin and Boyang Li and Chenguang Zheng and James Cheng and Fan Yu},
  title = {Seastar: Vertex-Centric Programming for Graph Neural Networks},
  booktitle = {EuroSys}, year = {2021}, doi = {10.1145/3447786.3456247},
  note = {\url{https://www.cse.cuhk.edu.hk/~jcheng/papers/seastar_eurosys21.pdf}}
}

@article{gala,
  author = {Damitha Lenadora and Nikhil Jayakumar and Chamika Sudusinghe and Charith Mendis},
  title = {{GALA}: A High Performance Graph Neural Network Acceleration LAnguage and Compiler},
  journal = {Proceedings of the ACM on Programming Languages}, volume = {9}, number = {OOPSLA2},
  year = {2025}, doi = {10.1145/3763113},
  note = {\url{https://charithmendis.com/assets/pdf/25-oopsla-gala.pdf}}
}

@misc{ebb,
  author = {Gilbert Louis Bernstein and Chinmayee Shah and Crystal Lemire and Zachary DeVito and Matthew Fisher and Philip Levis and Pat Hanrahan},
  title = {Ebb: A {DSL} for Physical Simulation on {CPUs} and {GPUs}},
  year = {2016}, howpublished = {arXiv:1506.07577v2},
  note = {\url{https://arxiv.org/abs/1506.07577v2}}
}

@article{simit,
  author = {Fredrik Kjolstad and Shoaib Kamil and Jonathan Ragan-Kelley and David I. W. Levin and Shinjiro Sueda and Desai Chen and Etienne Vouga and Danny M. Kaufman and Gurtej Kanwar and Wojciech Matusik and Saman Amarasinghe},
  title = {Simit: A Language for Physical Simulation}, journal = {ACM Transactions on Graphics},
  volume = {35}, number = {2}, pages = {20:1--20:21}, year = {2016}, doi = {10.1145/2866569},
  note = {\url{https://doi.org/10.1145/2866569}}
}

@article{graphit,
  author = {Yunming Zhang and Mengjiao Yang and Riyadh Baghdadi and Shoaib Kamil and Julian Shun and Saman Amarasinghe},
  title = {{GraphIt}: A High-Performance {DSL} for Graph Analytics}, year = {2018},
  journal = {Proceedings of the ACM on Programming Languages},
  volume = {2}, number = {OOPSLA},
  note = {\url{https://arxiv.org/abs/1805.00923}}
}

@article{taco,
  author = {Fredrik Kjolstad and Shoaib Kamil and Stephen Chou and David Lugato and Saman Amarasinghe},
  title = {The Tensor Algebra Compiler}, journal = {Proceedings of the ACM on Programming Languages},
  volume = {1}, number = {OOPSLA}, pages = {77:1--77:29}, year = {2017}, doi = {10.1145/3133901},
  note = {\url{https://tensor-compiler.org/files/kjolstad-oopsla17-tensor-compiler.pdf}}
}

@article{sparseformats,
  author = {Stephen Chou and Fredrik Kjolstad and Saman Amarasinghe},
  title = {Format Abstraction for Sparse Tensor Algebra Compilers}, year = {2018},
  journal = {Proceedings of the ACM on Programming Languages},
  volume = {2}, number = {OOPSLA}, doi = {10.1145/3276493},
  note = {\url{https://arxiv.org/abs/1804.10112}}
}

@inproceedings{sparsetir,
  author = {Zihao Ye and Ruihang Lai and Junru Shao and Tianqi Chen and Luis Ceze},
  title = {{SparseTIR}: Composable Abstractions for Sparse Compilation in Deep Learning},
  year = {2023},
  booktitle = {Proceedings of the 28th ACM International Conference on Architectural Support for Programming Languages and Operating Systems, Volume 3},
  doi = {10.1145/3582016.3582047},
  note = {\url{https://arxiv.org/abs/2207.04606}}
}

@inproceedings{graphchi,
  author = {Aapo Kyrola and Guy Blelloch and Carlos Guestrin},
  title = {{GraphChi}: Large-Scale Graph Computation on Just a {PC}},
  booktitle = {10th USENIX Symposium on Operating Systems Design and Implementation (OSDI)},
  year = {2012}, pages = {31--46},
  note = {\url{https://www.usenix.org/conference/osdi12/technical-sessions/presentation/kyrola}}
}

@inproceedings{xstream,
  author = {Amitabha Roy and Ivo Mihailovic and Willy Zwaenepoel},
  title = {{X-Stream}: Edge-centric Graph Processing using Streaming Partitions},
  booktitle = {ACM SOSP}, year = {2013},
  note = {\url{https://sigops.org/s/conferences/sosp/2013/papers/p472-roy.pdf}}
}

@inproceedings{legion,
  author = {Michael Bauer and Sean Treichler and Elliott Slaughter and Alex Aiken},
  title = {Legion: Expressing Locality and Independence with Logical Regions},
  booktitle = {International Conference for High Performance Computing, Networking, Storage and Analysis},
  year = {2012}, note = {\url{https://legion.stanford.edu/pdfs/sc2012.pdf}}
}

@inproceedings{mpnn,
  author = {Justin Gilmer and Samuel S. Schoenholz and Patrick F. Riley and Oriol Vinyals and George E. Dahl},
  title = {Neural Message Passing for Quantum Chemistry},
  booktitle = {Proceedings of the 34th International Conference on Machine Learning},
  series = {Proceedings of Machine Learning Research}, volume = {70},
  year = {2017}, pages = {1263--1272},
  note = {\url{https://proceedings.mlr.press/v70/gilmer17a.html}}
}

@inproceedings{pytorch,
  author = {Adam Paszke and Sam Gross and Francisco Massa and Adam Lerer and James Bradbury and Gregory Chanan and Trevor Killeen and Zeming Lin and Natalia Gimelshein and Luca Antiga and Alban Desmaison and Andreas Kopf and Edward Yang and Zachary DeVito and Martin Raison and Alykhan Tejani and Sasank Chilamkurthy and Benoit Steiner and Lu Fang and Junjie Bai and Soumith Chintala},
  title = {{PyTorch}: An Imperative Style, High-Performance Deep Learning Library},
  booktitle = {Advances in Neural Information Processing Systems},
  volume = {32}, year = {2019},
  note = {\url{https://proceedings.neurips.cc/paper/2019/hash/bdbca288fee7f92f2bfa9f7012727740-Abstract.html}}
}

@article{autodiff,
  author = {Atilim Gunes Baydin and Barak A. Pearlmutter and Alexey Andreyevich Radul and Jeffrey Mark Siskind},
  title = {Automatic Differentiation in Machine Learning: a Survey},
  journal = {Journal of Machine Learning Research},
  volume = {18}, number = {153}, pages = {1--43}, year = {2018},
  note = {\url{https://www.jmlr.org/papers/v18/17-468.html}}
}

@misc{stanfordbunny,
  author = {{Stanford University Computer Graphics Laboratory}},
  title = {The Stanford {3D} Scanning Repository: Stanford Bunny},
  year = {1994},
  note = {Zippered reconstruction; accessed September 21, 2026. \url{https://graphics.stanford.edu/data/3Dscanrep/}}
}
\endgroup
\end{document}